%% file: main.tex
\documentclass[doublespacing,11pt]{article}
\input{header}

\input{macros}

\begin{document}

\maketitle
\input{abstract}
\input{intro}
\input{background}
\input{saturated}

\include{simplecount}

\input{concldiss}
\clearpage
\newpage
\begin{singlespace}
\raggedright
\bibliographystyle{abbrvnat}
\bibliography{biblio}
\end{singlespace}

\end{document}

%% file: header.tex
\usepackage{setspace}
\usepackage{graphicx} 
\usepackage{times,mathptmx,courier}
\usepackage[scaled=.92]{helvet}
\usepackage{pifont}
\usepackage{checkend}	
\usepackage{graphicx}	
\usepackage[dvipsnames]{xcolor}

\usepackage{booktabs}
\usepackage{amssymb}
\usepackage{amsmath}
\usepackage{amsfonts}
\usepackage{amsthm}
\usepackage{listings}
\usepackage[printonlyused,nohyperlinks]{acronym}

\usepackage{color}
\definecolor{greytext}{gray}{0.5}
\usepackage{comment}

\usepackage{algorithm}
\usepackage{algpseudocode}

\usepackage[numbers,sort&compress]{natbib}

\usepackage[format=hang,indention=-1cm,labelfont={bf},margin=1em]{caption}
\usepackage{url}
\usepackage{hyperref}
    
\hypersetup{colorlinks,allcolors=NavyBlue
}

\title{Issues of parameterization and computation for posterior inference in partially identified models}
\author{Seren Lee \and Paul Gustafson}
\date{\small{Department of Statistics, University of British Columbia}}

%% file: macros.tex
\DeclareUrlCommand\DOI{}
\newcommand{\doi}[1]{\href{http://dx.doi.org/#1}{\DOI{doi:#1}}}



\newtheorem{example}{Example}



\DeclareMathOperator{\logit}{logit}
\DeclareMathOperator{\expit}{expit}

\usepackage [ margin =1.25 in , top =1.25 in , bottom =1.25 i n ]{ geometry}


%% file: abstract.tex
\begin{abstract}
A partially identified model, where the parameters can not be uniquely identified, often arises during statistical analysis. 
While researchers frequently use Bayesian inference to analyze the models, when Bayesian inference with an off-the-shelf MCMC sampling algorithm is applied to a partially identified model, the computational performance can be poor. It is found that using importance sampling with transparent reparameterization (TP) is one remedy. This method is preferable since the model is known to be rendered as identified with respect to the new parameterization, and at the same time, it may allow faster, i.i.d. Monte Carlo sampling by using conjugate convenience priors. 
In this paper, we explain the importance sampling method with the TP and a pseudo-TP. We introduce the pseudo-TP, an alternative to TP, since finding a TP is sometimes difficult. Then, we test the methods' performance in some scenarios and compare it to the performance of the off-the-shelf MCMC method - Gibbs sampling - applied in the original parameterization. While the importance sampling with TP (ISTP) shows generally better results than off-the-shelf MCMC methods, as seen in the compute time and trace plots, it is also seen that finding a TP which is necessary for the method may not be easy. On the other hand, the pseudo-TP method shows a mixed result and room for improvement since it relies on an approximation, which may not be adequate for a given model and dataset.
\end{abstract}

%% file: intro.tex
\section{Introduction}\label{sec:1:intro}
During research that uses Bayesian inference, researchers set the scientific problem, gather the data, learn from the data about fixed but unknown parameters they are interested in by treating the parameters as random variables, and conclude based on knowledge that distributions describe. This research style is widely used in many areas, including medical and biological science. As the parameters are treated as random variables, the information researchers get is the posterior distribution over the parameter space. Bayesian inference is instantiated by a simple but powerful rule based on Bayes theorem,
\begin{equation}\label{eqn:1:Bayes}Posterior \propto Prior \times Likelihood.\end{equation}
Sometimes, the resulting posterior distribution is in a pre-defined, known distribution family, such as the normal distribution or the beta distribution, which statisticians have already characterized. However, in many more cases, even finding a closed-form posterior density function is not feasible. An alternative is generating a Monte Carlo sample from the desired distribution.
This process is usually enabled by the family of sampling methods called Markov chain Monte Carlo (MCMC), which makes Bayesian inference more realizable. As MCMC doesn't require the closed form of the target cumulative distribution function, it can be used even if the posterior distribution is very complicated or high dimensional.

However, like almost all algorithms and methods in the real world, MCMC is imperfect. Ideally, chains (iterations) of a sample converge to the target distribution. But, MCMC can have slow convergence and mixingness, consequently increasing the running time of sampling and providing results that may not fully reflect the target posterior distribution. There are multiple scenarios of MCMC failure. One notable scenario is especially when the model parameters can't be fully and uniquely identified,  which is called the partially identified model
\citep{Gustafson-2015-Partial}.

Fortunately, this computational challenge of fitting a partially identified model may be relieved by applying a transparent parameterization and importance sampling to MCMC or Monte Carlo i.i.d sampling, if the posterior is simple enough \citep{Gustafson-2015-Partial, Pirikahu-2021-Bayesian}. These methods ease the calculation burden by sampling from the posterior distributions of transformed parameters. If this computational workflow is applicable, we can expect improvement in the performance of the sampling.

In this paper, necessary background, including the partially identified model, transparent reparameterization with importance sampling, and MCMC diagnostic tools, is explored (Section 2). Some scenarios of the partially identified model arising from the Generalized linear model (GLM) with nonignorable missing data are explored. Then, simulation results with traditional off-the-shelf MCMC and transparent reparameterization with importance weighting are discussed (Sections 3 and 4). While Sections 3 and 4 both cover the scenario of a partially identified model in the GLM, the problem in Section 3 involves binary variables only, and the problem in Section 4 involves a count variable. Even though Section 4 has a simpler structure, it turns out that transparent reparameterization (TP) is not available, and therefore, an alternative method is devised. Lastly, the advantages and challenges of the sampling algorithm with transparent reparameterization and importance weighting are discussed and extended to further research topics (Section 5).

%% file: background.tex

\section{Background}
\subsection{Partially identified model}\label{sec:2.1:pim}
The term \textit{partially identified model} refers to the identification of the model. We say a model is well identified when distinct parameter values cannot yield the same distribution of observable data \citep{Gustafson-2010-Bayesian, Gustafson-2015-Partial}. Conversely, if this property doesn't hold, we have a nonidentified model. 
This concept can be interpreted in a similar way to a solution of an equation system in mathematics: unique solution versus infinitely many solutions. When the equations are insufficient to yield a unique solution, there can be infinitely many solutions. Similarly, if the data are not rich enough for learning all parameters, the model can be nonidentified.
Even if the partially identified model is not different from the nonidentified model from a mathematical perspective, in practice, it is sitting somewhere between the fully identified model and the nonidentified model. Lack of identification can be an inevitable consequence of what is known about and what isn't known about the data-generating process, but this could still lead to some level of useful inference. This describes the important feature of the partially identified model that we distinguish from the nonidentified model.

When the model is partially identified, the performance of MCMC can be poor. In fact, performance can further worsen as the model deviates further from the fully identified model. 

A simple example of the identified model is linear regression on the data of $X$ and $Y$.
\begin{example}\label{ex:1:slr} Simple linear regression:
$$(Y \mid X) \sim N(\beta_0 + \beta_1 X,\sigma^2).$$
\end{example}
\noindent When the data consists of two variables $X$ and $Y$, researchers are typically interested in $\beta_1$.
This model is fully identified. On the other hand, if we tweak this example to accommodate missing entries, it can result in a partially identified model. This example is a variation of the example introduced in \citep{Gustafson-2022-summer}.
\begin{example}\label{ex:2:missingslr} Simple linear regression with missing entries:
$$(Y \mid X) \sim N(\beta_0 + \beta_1 X,\sigma^2),$$
$$(R \mid X,Y) \sim Bern(\expit(\gamma_0 + \gamma_1 X + \gamma_2 Y + \gamma_3 XY)).$$
\end{example}
\noindent where $R$ is binary variable so that $R=1$ means $Y$ is observed, and $R=0$ indicates that $Y$ is missing. This data structure involves the same variables $(X,Y)$ as above, but some proportion of the $Y$ entries are missing. Researchers are still interested in $\beta_1$ but are also forced to consider other parameters, notably $(\gamma_2, \gamma_3)$, which describe the relationship between the missingness of $Y$ and $Y$ itself.  In detail, $\gamma_2$ and $\gamma_3$ describe ignorable missing data \citep{little-2019-statistical}, with the special case of missing at random (MAR) corresponding to $\gamma_2 = \gamma_3 = 0$ in this case, versus nonignorable missing data \citep{little-2019-statistical} and missing not at random (MNAR), where $\gamma_2 \neq 0$ or $\gamma_3 \neq 0$. This change makes the model partially identified since $Y$ is used to explain $R$ while $R$ explains the missingness of $Y$.

\subsection{Transparent reparameterization (TP) and importance sampling}
\subsubsection{Transparent reparameterization}\label{sec:2.2.1:tp}
Transparent parameterization \citep{Gustafson-2015-Partial}, which is a mathematical tool for reshaping and converting parameters, results in a transformed set of parameters $(\phi, \lambda)$ derived from the original parameter set $\theta$ by thinning out, changing, and introducing new parameters. 
Let $h$ be the invertible function for reparameterizing one set into another. For the function $h$ to be a transparent reparameterizing function, two conditions are to be satisfied. Given the vector of original parameters, $\theta \in \mathbb{R}^k$, and the result of the reparameterization $h(\theta) = (\phi,\lambda)$ those conditions are as follows:

\begin{enumerate}
\item \textbf{Independence between $\lambda$ and the data}. 

Let $\pi(d_n \mid \phi, \lambda)$ be the conditional density function of the data given parameter. We assume $d_n$ only depends on $\phi$. Therefore, $\pi(d_n \mid \phi, \lambda) = \pi(d_n \mid \phi)$.
\item \textbf{Applicability of regular parametric asymptotic theory} 

$\sqrt{n}$-consistent estimation of $\phi$ obtains through $\pi(d_n \mid \phi)$, i.e., on estimator $\tilde{\phi}$ constructed from $\pi(d_n \mid \phi)$ satisfies
$\tilde{\phi} - \phi = O_p(1/\sqrt{n})$.
\end{enumerate}

\noindent When both conditions are satisfied, we say $h(\theta) = (\phi,\lambda)$ is a transparent reparameterization.

The main role of a transparent reparameterization (TP) is to partition the parameters into two parts by identification when the model is partially identified as in \autoref{sec:2.1:pim}. We only keep the parameters contributing to the likelihood while removing unnecessary parts to minimize the number of elements in the first part, $\phi$. Then, the second part, $\lambda$, can be chosen arbitrarily as long as it keeps the invertibility of $h$ and matches the total dimension in the original parameterization. Due to the arbitrariness of $\lambda$, there will be countless TPs, if a TP exists. The choice of $\lambda$ is left to the researchers. Sometimes, $\lambda$ can be chosen to ensure good sampling performance. Or, $\lambda$ can be a subset of the original parameters that researchers can readily interpret. 

Finally, there is a chance that TP does not exist. Then, pseudo-TP can be considered. In this paper, pseudo-TP is a reparameterization with a function $h$ that is close to TP but not exactly since 
$d_n$ and $\lambda$ is weakly dependent given $\phi$ so that $\pi(d_n \mid \phi, \lambda) \neq \pi (d_n \mid \phi)$, while TP has a necessary condition that $d_n \perp\!\!\!\perp \lambda \mid \phi$.

\subsubsection{Convenience priors and convenience  posteriors}
\label{sec:2.2.2:convenience}
When Bayesian analysis is instantiated, there are prior distributions specified by the researchers, ideally based on existing knowledge about the scientific context at hand. However, those distributions are not always convenient for calculation purposes in the Bayesian analysis. Indeed, there can be prior distributions that are more calculation-friendly; however, they may differ from those that researchers specify. As introduced in \citep{Gustafson-2015-Partial}, those convenient but possibly different prior distributions from those that researchers specify are called \emph{convenience prior distributions}.  Accordingly,  we call the posterior distribution derived from convenience prior distributions \emph{convenience posterior distributions}. In this paper, we place $^{*}$ next to convenience density functions to differentiate them from the desired density functions.

\subsubsection{Importance sampling}\label{sec:2:2:3:importance}
Importance sampling \citep{Kloek-1978-Bayesian, Glynn-1989-Importance}, is a widely applicable technique in computational statistics that helps find a given distribution's properties. The most noticeable feature of the importance sampling is that it can be applied to distributions which are not easy to draw a sample from.
It generally can be explained in one simple expression:
\begin{equation}\label{eqn:2:im}
W_i \propto \frac{f(X_i)}{g(X_i)},
\end{equation} 

\noindent where $X_i$ is the $i$-th observation in the sample $X_1, X_2, \dots, X_n$, f(X) is proportional to the original density of interest, g(X) is proportional a new, easy-to-sample density and $W_i$ is weight of the observation.
Rather than trying hard to draw a sample from the original density $f(X)$, we first draw an i.i.d. sample of $n$ observations from the new easy-to-sample density $g(X)$ and find desired properties of the original distribution by using this sample and the ratio in \autoref{eqn:2:im}. 
As the weight is a gap between actual density and something proportional to the density, weighting allows appropriate adjustment to the sample from $g(X)$. 
Consequently, extra processes other than sampling and weighting, such as explicit normalization of densities, are not required. The critical idea of importance sampling is changing the problem for the benefits of time and complexity while still linking back to the original problem, which can be easily and frequently found in reducing problems \citep{Garey-1974-Some} in Computer Science. We achieve the desired result with different methods for different problems as long as the reduction is well-designed. 

Specifically, importance sampling can be used to adjust for the discrepancy when using convenience distributions rather than their desired counterparts \citep{Gustafson-2015-Partial}.
Drawing a sample from the convenience posterior obtained using the TP and then applying importance sampling may look like extra steps compared to simply using  MCMC directly in the original parameterization. However, since this method draws a sample from the convenience posterior density $\pi^*(\phi \mid d_n)$ where $\phi$ is fully identified and the convenience prior density $\pi^*(\lambda \mid \phi)$, the computational efficiency is better than traditional MCMC methods sampling on the partially identified model that can't uniquely determine $\theta$.

The method, introduced in \citep{Gustafson-2015-Partial}, is summarized in \autoref{alg:1:im}. Given inputs of the user (researcher) specified prior density function $\pi_0(\theta)$ and convenience prior density function $\pi^{*}(\phi,\lambda) = \pi^{*}(\lambda \mid \phi) \cdot \pi^{*}(\phi)$, this algorithm returns a posterior sample of $\theta$. Note that $\pi(\phi, \lambda)$ is required, but since the researchers specify the prior $\pi_0(\theta)$, this density will be transformed using $\pi(\phi,\lambda) = \pi_0(h^{-1} (\phi,\lambda)) \cdot \mid \nabla h^{-1} (\phi,\lambda) \mid$ \citep{Ross-2010-First}.

\begin{algorithm}
\caption{Importance sampling for TP (ISTP)}\label{alg:1:im}
\textbf{Input} \\  Prior density function (user-specified) $\pi(\phi,\lambda)$ \\ 
Convenience prior distribution $\pi^{*}(\phi,\lambda) = \pi^{*}(\lambda \mid \phi) \cdot \pi^{*}(\phi)$\\
\textbf{Output} \\
Sample from desired posterior distribution $\pi(\theta \mid d_n)$
  \begin{algorithmic}[1]
  \State Calculate the marginal convenience posterior density function of $\phi$, $\pi^{*} (\phi \mid d_n)$.
  \State Find the joint convenience posterior distribution, $\pi^{*}(\phi, \lambda \mid d_n) = \pi^{*} (\phi \mid d_n) \pi^{*}(\lambda \mid \phi)$. 
  \If {$\pi^{*}(\phi, \lambda \mid d_n)$ is composed of common distributions
  available in \texttt{R}}
  \State Generate $m$ draws composed of i.i.d. Monte Carlo draws using \texttt{R}.
  \Else
  \State Generate $m$ draws using an off-the-shelf MCMC package.
  \EndIf
  \State
  Let the $i$-th draw be $(\phi_i, \lambda_i)$.
  \State Calculate weight $W_i = \frac{\pi(\phi_i, \lambda_i)}{\pi^{*}(\lambda_i \mid \phi_i) \cdot \pi^{*}(\phi_i)}$.
  \State The final weight $w_i$ is normalized, $w_i = \frac{W_i}{\sum_{j=1}^m{W_j}}$.
  \State Resample the $(\phi_i, \lambda_i)$ draws using weights $w_i$ to get a sample from $\pi(\phi,\lambda \mid d_n)$
  \State Return a final sample mapped back into the parameterization of $\theta$.
  \end{algorithmic}
\end{algorithm}
\subsection{MCMC diagnostic tools}

\subsubsection{Effective Sample Size (ESS)}
Effective sample size (ESS) \citep{Kish-1965-Survey} refers to the adjusted sample size that reflects the amount of reliable information in a Monte Carlo sample. ESS is inversely proportional to autocorrelation in an MCMC sample. An autocorrelated sample with $ESS = n$ contains the same amount of information about the underlying distribution as a hypothetical i.i.d sample of size $n$. ESS may be much smaller than the actual sample size, and small ESS implies that MCMC chains aren't mixing well, which is not an ideal performance of MCMC.

For importance sampling, the common ESS of all parameters is given by the formula \citep{Gustafson-2015-Partial, Glynn-1989-Importance, Doucet-2001-Sequential}
\begin{equation}
    ESS = 1/ \left(\sum_{i=1}^m w_i^2\right).
\end{equation}
For the MCMC sampling using STAN \citep{R-rstan} or JAGS \citep{R-rjags}, the estimation of ESS of each parameter can be calculated based on autocorrelation of the posterior sample, as implemented by bespoke functions in packages \citep{R-rstan, R-rjags, R-coda}. If importance weighting is applied {\em after} an MCMC sampling package is used, the final estimated ESS is calculated by multiplying losses of information from each step:

\begin{equation}
    ESS = \left(1/ \left(\sum_{i=1}^m w_i^2\right)\right) \times ESS_p / m,
\end{equation}
where $m$ is the actual sample size and $ESS_{p}$ is the estimated ESS from the package.

\subsubsection{Multivariate ESS}\label{sec:2.3.2:multiESS}
When MCMC is used for posterior sampling, the ESS covered in the last section pertains to a single parameter. We can indeed report the ESS for the parameter of most interest, to diagnose MCMC performance. However, in models with multiple parameters, especially high-dimensional, this single ESS may not be representative of overall MCMC performance. ESS estimates of some parameters can be indeed small, while others are closer to the actual sample size. Since the posterior distribution is inherently a joint distribution of all parameters, ideally, multiple parameters should be considered.
As this issue arises from using a single ESS, it can be alleviated by using an estimate of ESS for multivariate parameters. The multivariate ESS estimation introduced in \citep{vats2015multivariate}, is based on 
$$ESS = m\cdot\frac{|\Lambda|^{\frac{1}{p}}}{|\Sigma|^{\frac{1}{p}}},$$ where $m$ is the actual sample size, $p$ is the number of parameters, $\Lambda$ is the sample covariance matrix, and $\hat\Sigma$ is the estimation of a Monte Carlo standard error. The matrix $\Sigma$ is positive definite and defined under Markov chain central limit theorem (CLT) \citep{jones-2004-markov}, $\sqrt{n} (\hat{f}_n - E(f(X_i))) \rightarrow^d (0,\Sigma)$ where $X_1, X_2, \dots, X_n$ is Markov chain with invariant probability distribution equal to the target distribution and $\hat{f}_n = \frac{1}{n}\sum_{i=1}^n f(X_i)$. Since we would like to have the Monte Carlo standard error of all parameters, $f(X)$ is the identity function.
All simulations in this paper use the R function \texttt{multiESS()} in \texttt{mcmcse} package \citep{R-mcmcse}. 
\subsubsection{Trace plot}
A trace plot is a graph showing the sample's evolution versus the state (sample) number. Similarly to ESS, the trace plot can show the convergence and autocorrelation. 
\begin{figure}
    \centering
    \includegraphics[width = 1\textwidth]{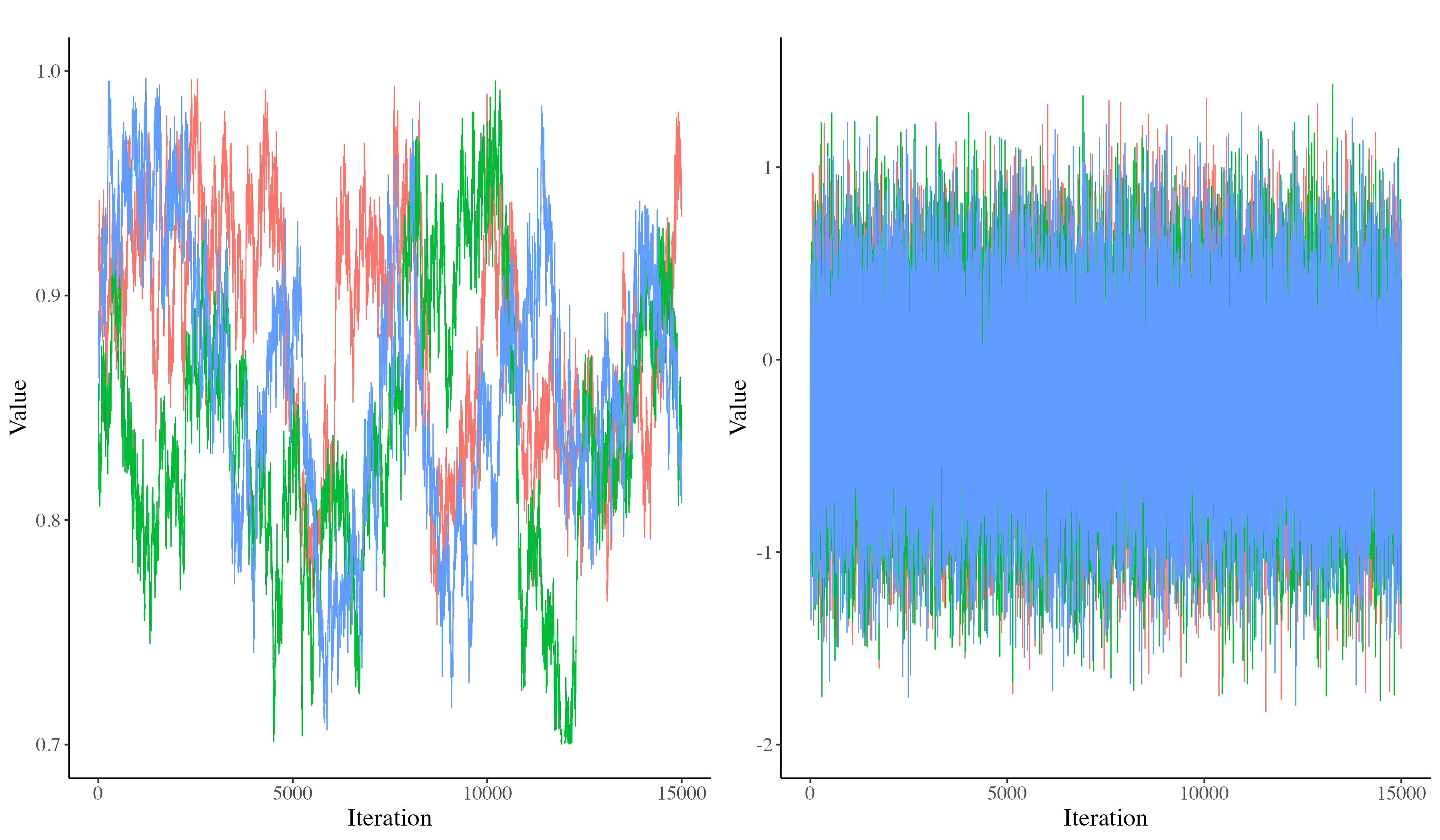}
    \caption{Two trace plots}
    In both plots, there are $3$ chains of $15000$ iterations after $2000$ burn-ins, respectively.
    \label{fig:1:exampleTr}
\end{figure}
If there is a prominent shape like the left plot in \autoref{fig:1:exampleTr}, a sample is less believable since they are highly autocorrelated. Also, three chains don't agree with each other. On the contrary, the right plot in \autoref{fig:1:exampleTr} doesn't have any prominent shape; the three chains are overlapped, and mixing well. Even though there is no standard rule for determining the performance of MCMC through trace plots and there is no way to consider multiple parameters simultaneously in a single plot, it is worth using them since they are very straightforward to compare and interpret.

%% file: saturated.tex
\section{Logistic regression on multiple variables with missing entries in the dependent variable}\label{sec:3:sat}
In this section, an example of a partially identified model is introduced, a corresponding TP is derived, and simulation results with traditional MCMC and the ISTP algorithm on multiple scenarios of poor MCMC performance are summarized.

\newcommand{\xdes}{\mathbf{X}_{des}}
\autoref{ex:3:logisticregsat} is an extension and variation of the mock data example introduced in unpublished lecture notes \cite{Gustafson-2022-summer}. The data are comprised of $n$ observations of $X$, a vector of independent variables with $p$ binary entries $X_1, X_2, \dots, X_p$, and a single dependent binary variable $Y$. Researchers are interested in the association between $X$ and $Y$. The challenging aspect of these data is
 that some entries of $Y$ are unavailable.  At this point, it is not yet clear whether a model 
 of $Y$ versus $X$ results in a partially identified model or an identified model. 
If $Y$ is suspected to be missing not at random (MNAR) i.e, the missingness of $Y$ is influenced by both $Y$ and $X$, and consequently the model is partially identified. The intuition behind this is the same as in \autoref{ex:2:missingslr} - Missingness of $Y$ is again explained by $Y$. After introducing a new binary variable $R$, which is 1 when $Y$ is observed and 0 when $Y$ is missing, the model researchers have designed can be summarized as below:
\begin{example}\label{ex:3:logisticregsat} Saturated logistic regression with incomplete outcome data:
\end{example}
\begin{itemize}
    \item \textbf{Data}\\
    \setstretch{1.6}
$\mathbf{X}  = (X_1, \dots, X_p)^T$\\
A design matrix of $\mathbf{X}$ is introduced to describe a saturated model. Let a vector $\xdes \in \mathbb{R}^{2^p}$ be a generic row of the design matrix, then\\
$\xdes  = (X_{0\dots0},\dots,X_{1,\dots1}) \text{ where }  X_{i_{1},i_{2},\dots ,i_{p}} = 1 \Leftrightarrow X_{1} = i_1, X_2 = i_{2}, \dots, X_p = i_p$\\
$\mathbf{X}_{des}$ has $2^p$ element where each element represent a combination of $X_1$ to $X_p$. As these elements cover all the combinations of $\mathbf{X}$, exactly one element is 1, and the others are zero.
$Y$ is the
dependent variable, which is binary.\\
$R$ is also a binary variable that describes the missingness of $Y$. $R = 1$ if and only if $Y$ is observed, and $R=0$ if and only if $Y$ is not available.

\newcommand{\bbR}{\mathbb{R}}

\item \textbf{Parameters}\\
\setstretch{2.0}
$\alpha  = (p_{0\dots0}, \dots,p_{1\dots1})^T$  where $\alpha$ belongs to the probability simplex on $2^p$ categories and  \\
$~~~~~~p_{i_{1},i_{2},\dots, i_{p}}  = Pr(X_{1} = i_1, X_2 = i_{2}, \dots, X_p = i_p)$; \\
$\beta  \in \mathbb{R}^{2^p}$;\\
$\gamma \in \mathbb{R}^{2^p};\\
\delta \in \mathbb{R}^{2^p}.$

\item \textbf{Model - Multinomial and Logistic regression}\\
$\mathbf{X}_{des}  \sim Multinom(\alpha,1)$\\
A multinomial distribution is suitable as the sum of each row is fixed to 1, and the sum of the probability of each column being one is 1 while those columns are independent of each other.\\
$(Y \mid \mathbf{X}_{des})  \sim Bern(\expit(\mathbf{X}_{des} {\beta}));\\
(R \mid \mathbf{X}_{des}, Y = 0)
 \sim Bern(\expit(\mathbf{X}_{des} \gamma));\\
(R \mid \mathbf{X}_{des}, Y = 1)
 \sim Bern(\expit(\mathbf{X}_{des} (\expit(\logit(\gamma+\delta))))).$\\
Since both $Y$ and $R$ are binary, $Y$ given $X$ and $R$ given $(X, Y)$ are explained with the logit scale of saturated linear regression in this model.
\end{itemize}

\noindent In summary, the unknown parameters are $\alpha, \beta, \gamma, \delta$.Then,
uninformative priors are chosen for these parameters:
\begin{equation}\label{eqn:5:prior}
\begin{aligned}
        \alpha & \sim Dir(\mathbf{1}_{2^p}), &  \\
        \beta_{i} & \sim Unif(0,1) & (i = 0, \dots ,2^p-1), \\
        \gamma_{i} & \sim Unif(0,1) & (i = 0, \dots ,2^p-1), \\
    \delta_{i} & \sim Normal(0, \sigma^2) &  (i = 0, \dots ,2^p-1). 
\end{aligned}
\end{equation}
Since elements of $\delta$ are log odds ratio of $(R, Y)$ given $X$, the magnitude of $\delta$ describes the departure of missingness from MAR and, therefore, the distance from a well-identified model.
Researchers set $\sigma^2$ based on their belief about how much deviation from MAR could exist. Consequently, in this paper, the prior variance of $\delta_i$, $\sigma^2$, is the hyperparameter varied in the simulation to test the impact of MNAR.


\newcommand{\supf}{^{(0)}}

\subsection{Transparent reparameterization}
If we directly apply a standard MCMC sampling method, especially Gibbs sampling, with parameters $\alpha, \beta, \gamma, \delta$, the computational performance is poor since the elements of $\alpha$ are not independent of each other. The reason is that $\alpha$ belongs to the simplex where the sum of all elements of $\alpha$ is 1. This issue is simply resolved by keeping independent elements in parameters and performing MCMC sampling. In this paper, whenever quantity $x$ is constrained to have elements summing to one, we let $x\supf$ be the subset of $x$ without the first element. Then,
the parameter set, $\theta$, consists of $(\alpha\supf, \beta, \gamma, \delta)$. The dimension of $\theta$ is $dim(\theta)= 2^p - 1 + 3 \times 2^p   = 2^{p+2} - 1$. This dimension should be preserved after applying TP.
The issue with these parameters as given by $\theta$ is that some of them can't be directly learned from the data. Considering $Y$ is not available when $R=0$, while $X$ is always available,
we can directly learn the following parameters from data,

\begin{equation}
    \label{eqn:6:phi}
\begin{aligned}
\epsilon & = Pr(R=1) , \\
\zeta\supf & = Pr\left(\xdes\supf \mid R=0\right),  \\
         \eta\supf  & = Pr\left(\xdes\supf \mid R  = 1\right),  \\
\xi &= Pr(Y  = 1 \mid \xdes, R = 1). 
    \end{aligned}
\end{equation}
These parameters are taken to comprise $\phi$. The number of parameters in $\xi$ is the number of entries in $\xdes$, $2^p$, while $\zeta\supf$ and $\eta\supf$ have one less. Therefore, the dimension of $\phi$ is $dim(\phi) = 1 + 2^p + 2 (2^p-1) = 3 \times 2^p - 1$. Now, we are choosing $\lambda$ that makes $h(\theta) = (\phi,\lambda)$ invertible and preserves the dimension. The vector of log odds ratios $\delta$ is one option, while infinitely many $\lambda$ options are available as long as they meet the condition.
In the case of using $\lambda= \delta$, the function $h(\theta)$ is given by the system of linear equations below:
\begin{equation}\label{eqn:7:function_h}
\begin{aligned}
        \epsilon & = Pr(R = 1) = \sum_{x \in \xdes} Pr(R = 1 \mid x) \cdot Pr(x) \\ & = \sum_{x \in \xdes} \{Pr(R = 1 \mid x, Y= 1) \cdot Pr(Y = 1 \mid x) + (Pr(R = 1 \mid x, Y= 0) \cdot Pr(Y = 0 \mid x)  \} \cdot Pr(x)  \\ & = <\alpha, \gamma^*, \beta> + <\alpha, \gamma, \bar{\beta}>, \\
        \zeta & = Pr(\xdes \mid R =0) = Pr(\xdes, R =0) / Pr(R = 0)\\   
         & = \{Pr(\xdes, Y = 1, R =0) +Pr(\xdes, Y = 0, R =0) \}/ Pr(R = 0) \\ 
        & = \{\beta \cdot \alpha \cdot (1-\gamma^*) + (1-\beta) \cdot \alpha \cdot (1-\gamma)\} / (1-\epsilon), \\
        \eta & = Pr(\xdes \mid R =1) = Pr(\xdes, R =1) / Pr(R = 1)\\   
         & = \{Pr(\xdes, Y = 1, R =1) +Pr(\xdes, Y = 0, R =1) \}/ Pr(R = 1) \\ 
        & = \{\beta \cdot \alpha \cdot \gamma^* + (1-\beta) \cdot \alpha \cdot \gamma\} / \epsilon, \\
        \xi &=  Pr(Y \mid \xdes, R=1) = Pr(Y , \xdes, R=1) / Pr(\xdes, R=1)\\
        & = (\beta \cdot \alpha \cdot \gamma^*)/(\eta\cdot \epsilon),
    \end{aligned}
\end{equation}
where $<x,y,z>$ is the sum of an element-wise product of vectors, $<x,y,z> = \sum_{i=1}^{dim(x)} x_i \cdot y_i \cdot z_i$.

\noindent Then, the function $h^{-1}(\phi,\lambda)$ is given by the system of linear equations below:
\begin{equation}\label{eqn:8:function_h_inv}
    \begin{aligned}
  \alpha & =  Pr(\xdes) =  Pr(R = 1) \cdot P(\xdes \mid R = 1) + Pr(R=0) \cdot P(\xdes \mid R = 0)\\
  & = \epsilon \cdot \eta + (1-\epsilon) \cdot \zeta,\\
  \beta & = Pr(Y \mid \xdes) = Pr(Y, \xdes) / Pr(\xdes) = (Pr(Y, \xdes, R = 0) + Pr(Y, \xdes, R = 1))/\alpha\\
  & = \{\expit(\logit(\xi) - \delta) \cdot (1-\epsilon) \cdot \zeta + \xi \cdot \epsilon \cdot \eta\}/\alpha, \\
  \gamma &= Pr(R = 1 \mid \xdes, Y = 0) = Pr(R = 1, \xdes, Y = 0) / Pr(\xdes, Y = 0) \\ 
  &= (1 - \xi) \cdot \eta \cdot \epsilon / \{(1-\beta)\cdot \alpha\}.
    \end{aligned}
\end{equation}
\subsection{Simulation methods and results}\label{sec:3.2.:satsim}
To compare the performance of traditional MCMC and ISTP, we set up three scenarios that may differently burden the computation and sampling scheme. While arguments of sampling, $n, p, \sigma$, are varied respectively in each scenario, all other non-varied arguments are fixed to common values of $n = 3000$, $p = 3$ and $\sigma = 0.5$. Mock data is generated with the model specified in \autoref{ex:3:logisticregsat} by using parameter values produced with random generator functions in \texttt{R}. To reproduce these parameter values, a seed value of random generator functions is fixed through the simulation.
\begin{enumerate}
    \item The number of observations, $n$. (10 levels, 3 - 10000)

    Having enough information is essential in estimating; however, MCMC sampling can take too long or break down if the data is too large.  Therefore, mock data with a tiny number of observations $(n = 3)$ to many observations ($n=10000$) are tested while other conditions remain the same.
    \item The number of variables in $X$, $p$. (5 levels, 1 - 5)

    Estimating is generally harder if we have many parameters to estimate. Since the model is saturated, the number of total parameters increases exponentially in $p$.
    
    \item The deviation a prior allowed from the MAR assumption, as represented by the prior variance of $\delta_i$, $\sigma^2$. (7 levels, 0.01 - 10)

    This is related to how much the model is partially identified. With the wider prior, the missingness of $Y$ is permitted to deviate further from MAR. So, we expect that it will make the performance worse. In the simulation, a narrow prior ($\sigma = 0.01$) to a wide prior ($\sigma = 10$) are used.
\end{enumerate}

Two methods are tested with these three scenarios for generating a posterior sample of parameters $\theta = (\alpha, \beta, \gamma, \delta)$. 
For both methods, $45000$ draws are found for each parameter while the first method, JAGS, has three chains of 15000 draws each after 2000 burn-ins.

First, traditional MCMC is implemented with JAGS called from R \citep{R-rjags} with model and prior specification declared in terms of the original parameterization in \autoref{ex:3:logisticregsat} and \autoref{eqn:5:prior}. Three chains are used to verify that chains are mixing well.  In this method, JAGS draws a sample from the posterior distribution of parameters and missing data $Y$ given observed data $Y$ and $X$. 

In contrast, the second method uses ISTP and draws an i.i.d. Monte Carlo sample of parameters in $(\phi, \lambda)$ given observed data by using \texttt{R} base functions such as \texttt{rnorm()} and \texttt{rbeta()}. Since $\phi$ is newly introduced, convenience priors are required to be set. For these parameters and data structure, conjugate priors are available.
For $\epsilon$, prior $beta(1,1)$ will result in posterior of $beta(1 + n_{R=1}, 1+ n_{R=0})$. For other parameters, a Dirichlet prior $Dir((1,\dots,1))$ will result in posterior of $Dir((1 + n_{1}, \dots, 1+ n_{k}))$ where $k$ is the number of parameters and $n_{i}$ is the number of observations that belongs to condition (category) $i$. When finding weights for resampling, the ratio between two prior densities, true and convenience, is evaluated. Since a sample is drawn from the convenience posteriors, a change of variables \citep{Ross-2010-First} is required. Consequently, the determinant of the Jacobian of the inverse function $h^{-1}(\phi,\lambda)$ is found. In all simulations in this paper, the R function \texttt{jacobian()} in \texttt{R} package \texttt{pracma} \citep{R-pracma} is used. This \texttt{R} package and function enable the \texttt{R} user to access the MATLAB function \texttt{jacobian()} \citep{matlab-jacobian} for finding Jacobian matrix by using provided inputs of symbolic \texttt{R} functions that take parameters and density functions as functional arguments. Given these inputs, \texttt{jacobian()} function returns the Jacobian matrix at given data entries, which is used to find the determinant with another R function \texttt{det()}.

\large{\noindent\textbf{Results}}\\
While we also consider multiESS and median ESS across all elements of $\theta$ to reflect the performance of MCMC, we focus on the ESS and trace plot of a presumed target parameter $\beta_1 =  \logit Pr(Y = 1 \mid X_1 = 1, X_2 = 0, \dots, X_p = 0)$.

\noindent\textbf{Scenario 1 ($n = 3 - 10000 ~ (10 ~ levels), p = 3, \sigma = 0.5$)}

\begin{figure}[h]
    \centering
    \includegraphics[width = 0.7\textwidth]{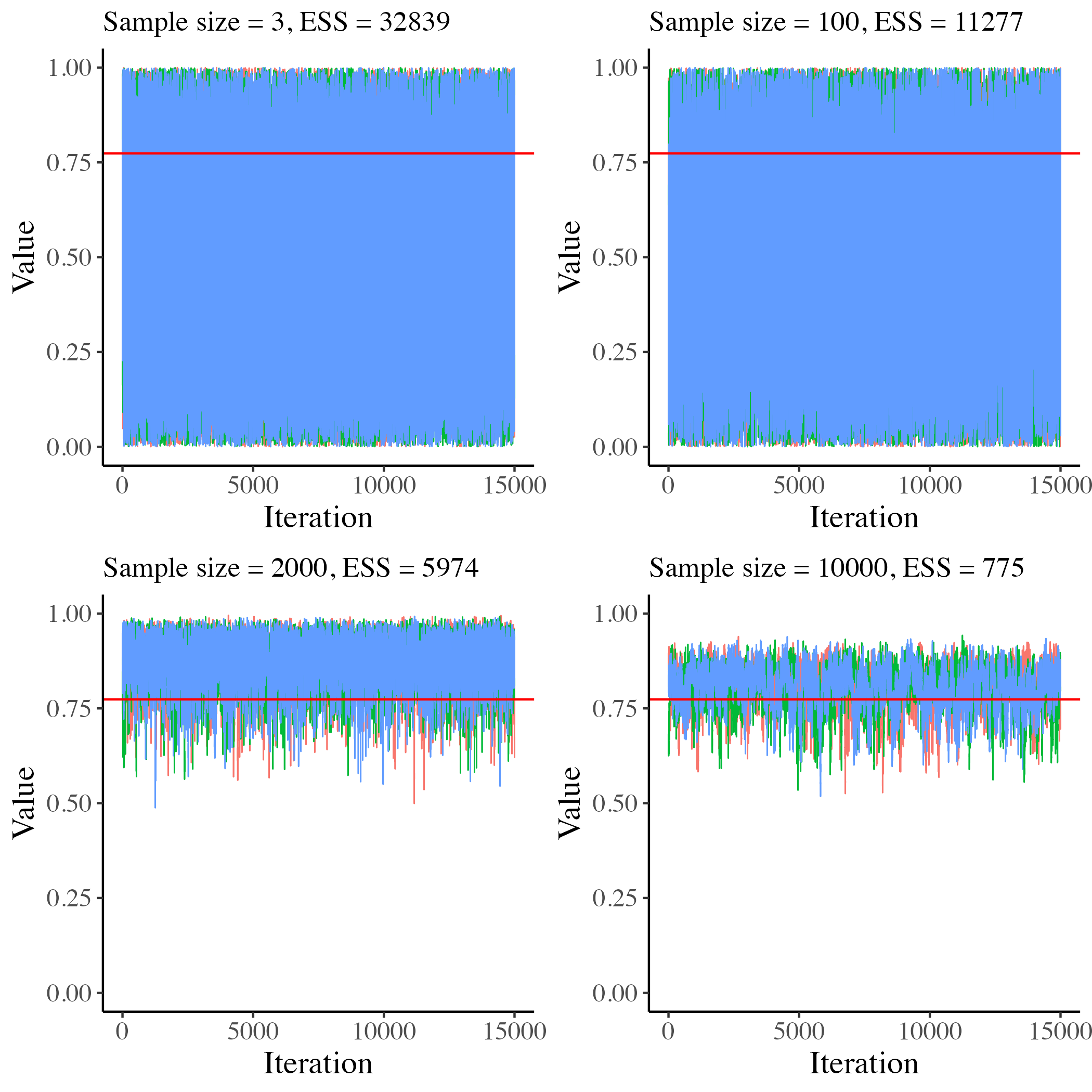}
    \caption{Method 1 (JAGS) Scenario 1 (The number of observations) - Trace plot of $\beta_1$}
    \label{fig:2:sat_n_tr}
        Trace plots and ESS of $\beta_1$. The red line denotes the true value of $\beta_1$. Among $10$ values of $n$, $n = 3,100,2000,$ and $10000$ are plotted.
\end{figure}

\begin{figure}
    \centering
    \includegraphics[width = 0.49\textwidth]{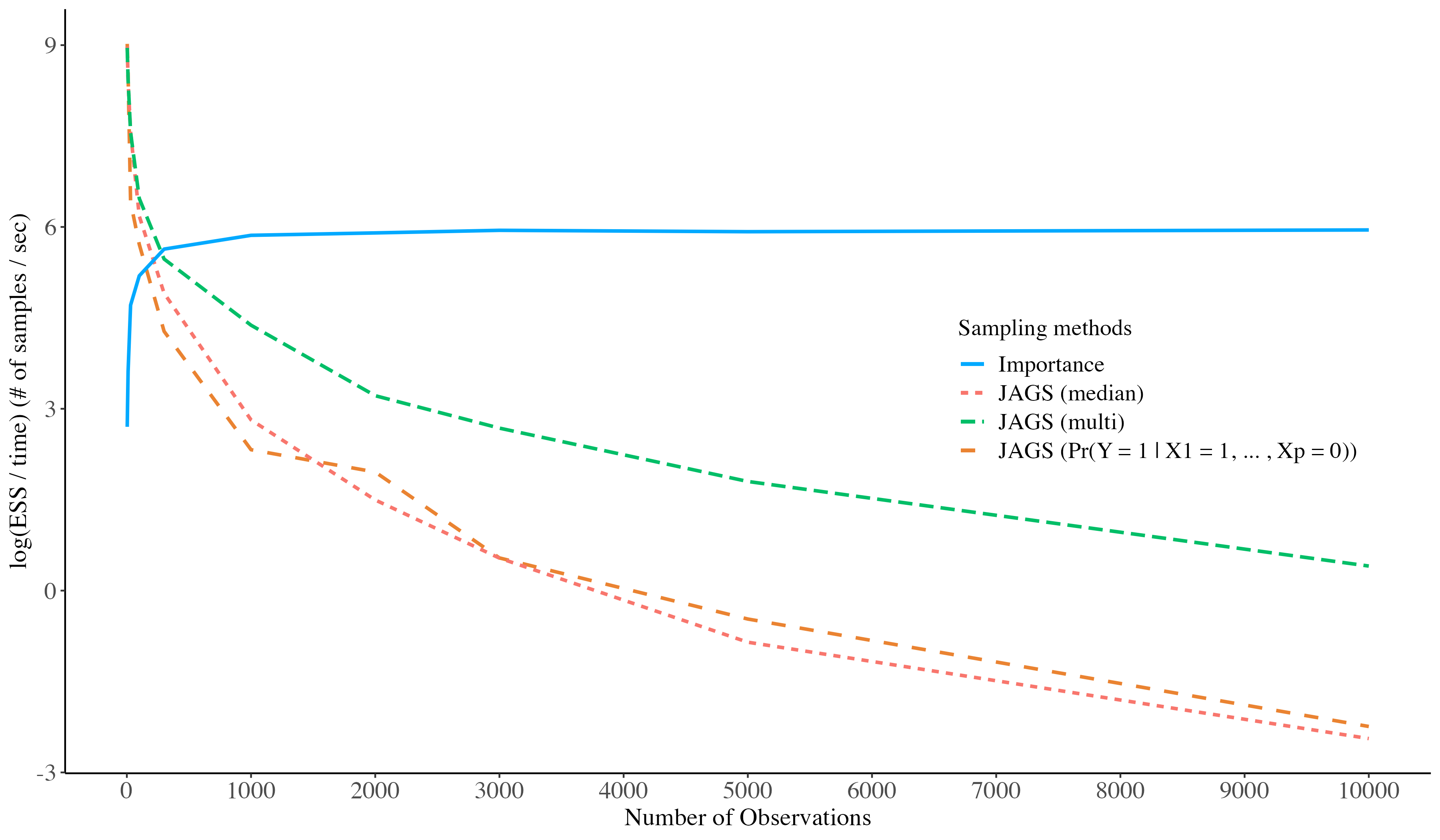}
    \includegraphics[width = 0.49\textwidth]{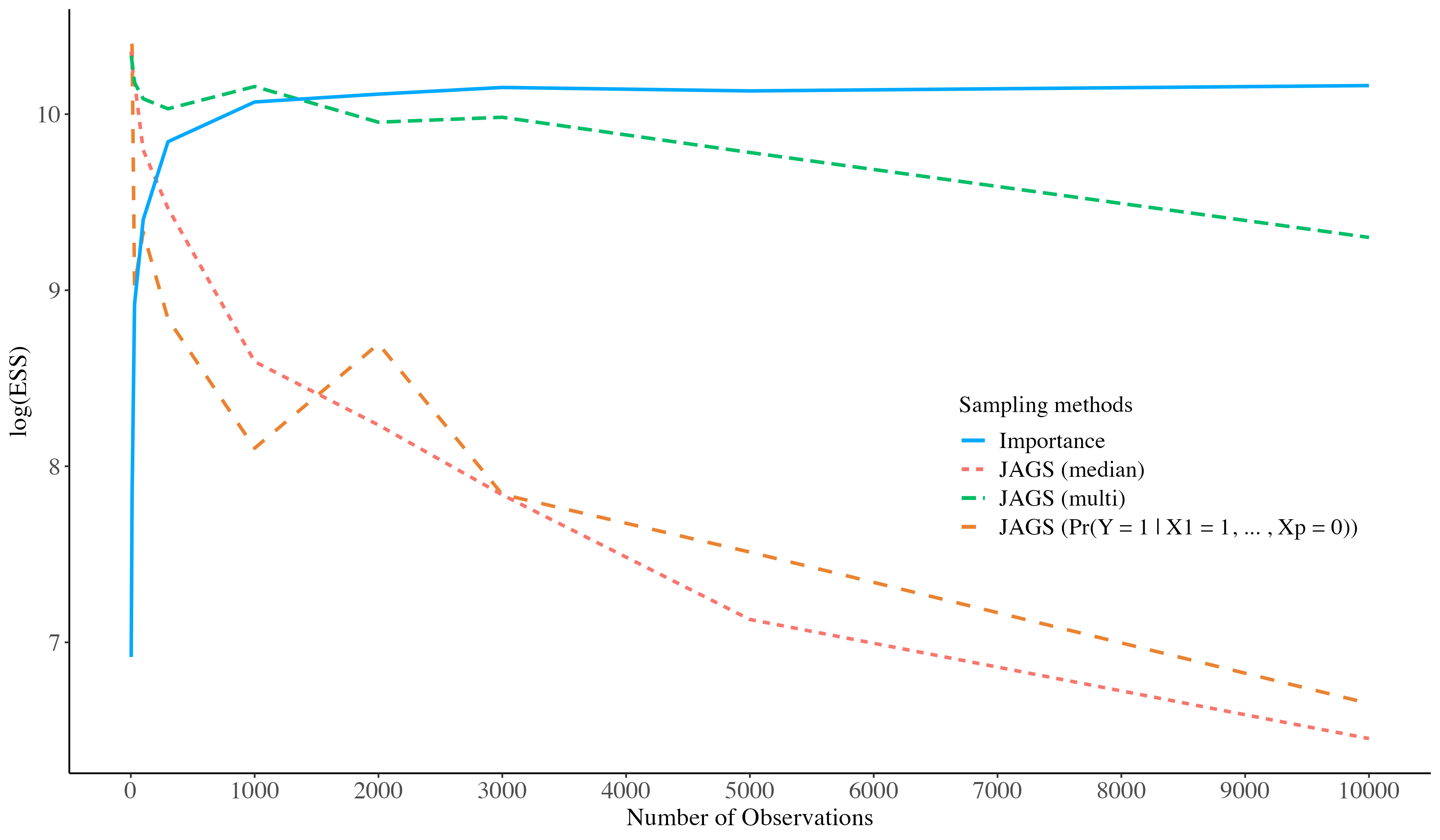}
    \includegraphics[width = 0.49\textwidth]{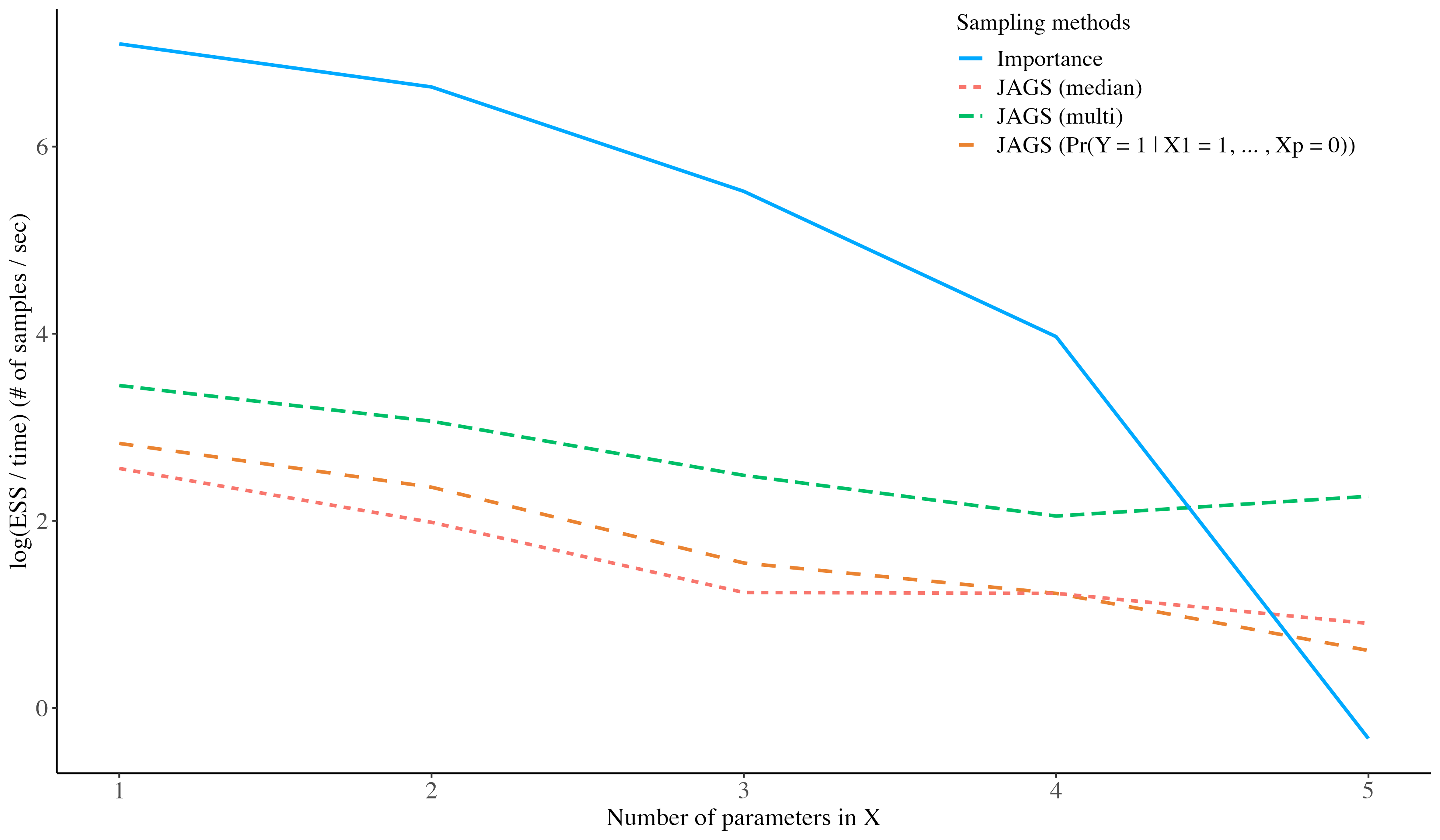}
    \includegraphics[width = 0.49\textwidth]{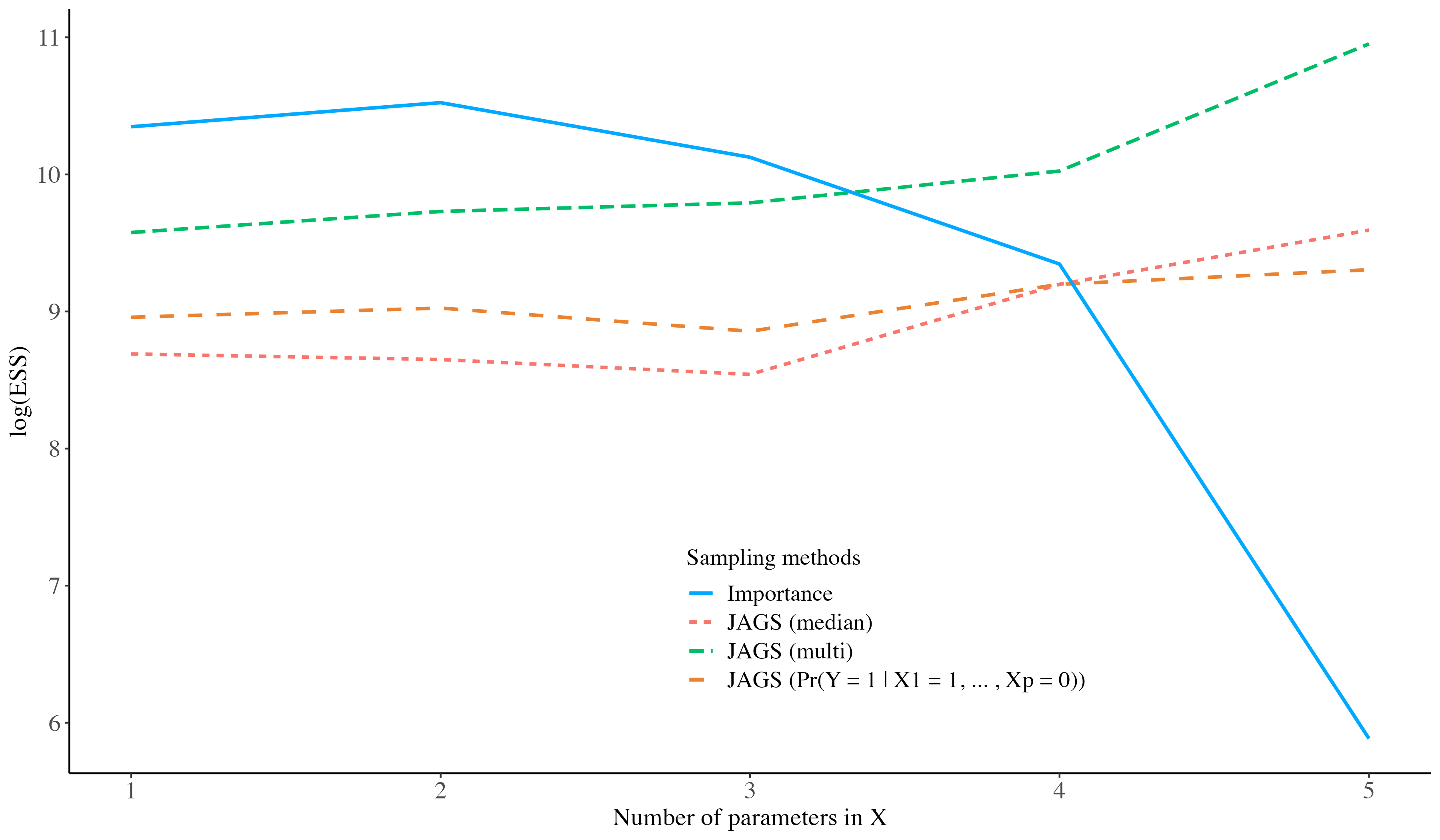}
    \includegraphics[width = 0.49\textwidth]{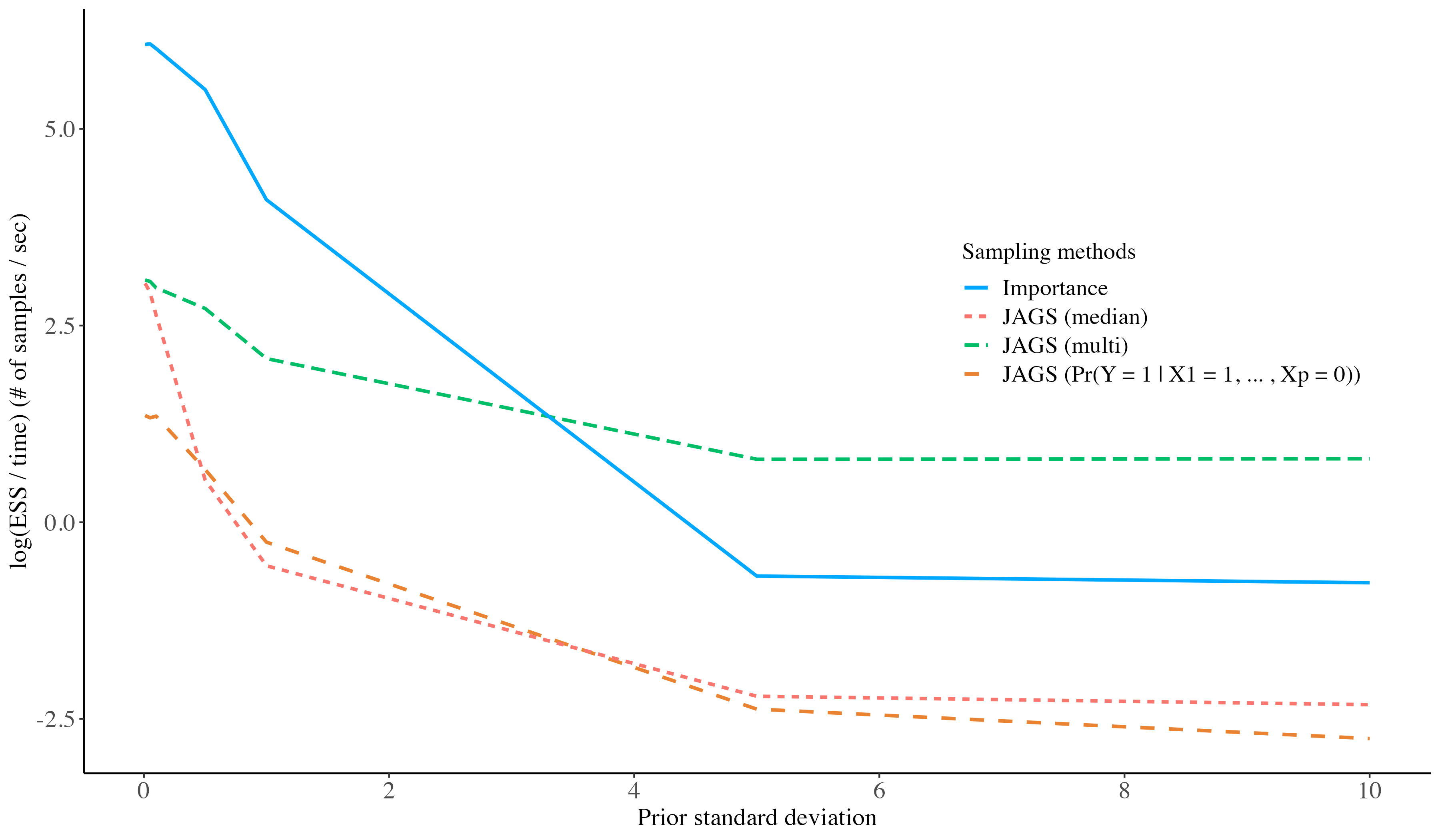}
    \includegraphics[width = 0.49\textwidth]{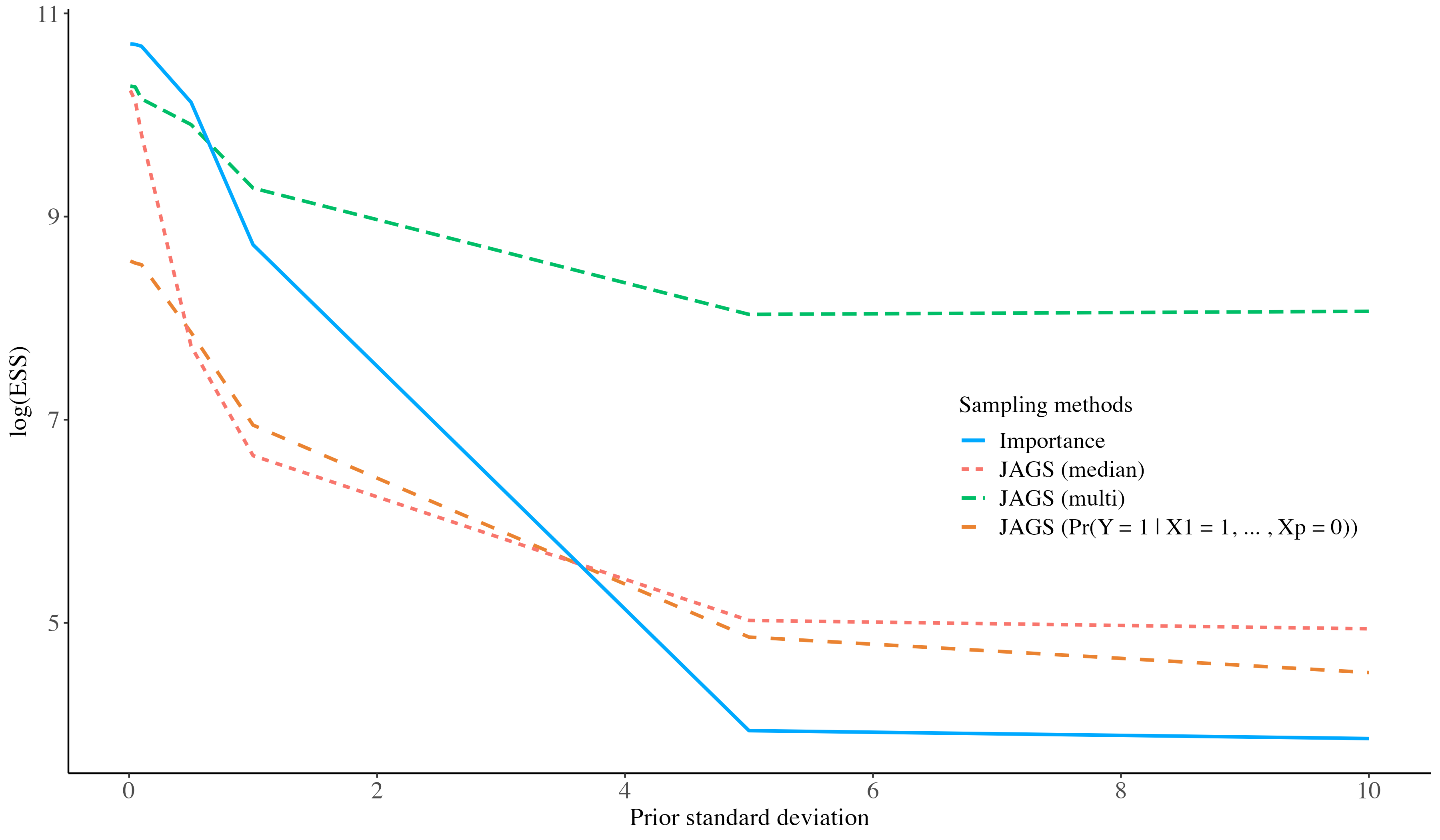}
    \caption{ESS and time in each method and scenario}
    \label{fig:5:saturated_ess_time}
    The left plots are ESS/time, and the right plots are ESS in each scenario and method. Method 1 has three lines each, the red line is with a median of ESS of all parameters, the green line is multiESS \citep{vats2015multivariate, R-mcmcse} in \autoref{sec:2.3.2:multiESS}, and the orange line is ESS of $\beta_1 = \logit Pr(Y = 1 \mid X_1 = 1, X_2 = 0, \dots, X_p = 0)$. One blue line represents the result of method 2. 
\end{figure}

    In \autoref{fig:2:sat_n_tr} that shows the performance of method 1, two things are notable - when $n$ is larger, the trace plot of $\beta_1$ has more apparent autocorrelation and the ESS of $\beta_1$ decreases. Considering that the posterior sample size is $N = 45000$, an ESS of $770$ is small enough to indicate that MCMC is not working well. Also, in the top left plot of \autoref{fig:5:saturated_ess_time}, the ESS/time of method 1 is good in a small sample size regardless of how ESS is found. However, it steeply decreases right after that. The main reason for this is the sampling time. ESS itself is not as steeply decreasing as ESS/time. 
    We also see that the sampling time increases exponentially from 4 seconds to 2 hours in \autoref{table:1:sat_n_time}.
    On the other hand, method 2 shows small ESS/time for the small sample size but has generally stable ESS/time and is better than method 1 when the sample size is large. And as in \autoref{table:1:sat_n_time}, time doesn't change much from one minute, and it is the same for ESS of $25000$ after $n = 1000$. And most of all, the time is much smaller than that of method 2.

    In conclusion, the performance of method 1 worsens with larger $n$ in terms of both ESS and time. In contrast, the performance of method 2 is worse than method 1 when the data size is small, but it gets better when there is enough data, and the performance does not change much in both time and ESS after that. Therefore, it shows better performance.

\begin{table}[h]
\centering
\begin{tabular}{c|rrrrrrrrrr}
  \hline
n & 3 & 10 & 30 & 100 & 300 & 1000 & 2000 & 3000 & 5000 & 10000 \\ 
\hline
  Method1 - JAGS  & 4 & 6 & 14 & 37 & 96 & 323 & 844 & 1483 & 2927 & 7292 \\ 
  Method2 - IM  & 68 & 67 & 67 & 67 & 67 & 67 & 68 & 67 & 67 & 67 \\ 
   \hline
\end{tabular}
\caption{Simulation time of methods 1 and 2 (sec) as $n$ increases} 
\label{table:1:sat_n_time}
Method 1 uses JAGS and draws a sample  from $\pi(\theta, y_{mis} \mid d_n)$. Method 2 uses ISTP and draws a sample from $\pi^*(\phi,\lambda \mid d_n)$.
\end{table}

\noindent\textbf{Scenario 2 ($p = 1 - 5 ~ (5 ~levels), n = 3000, \sigma = 0.5$)}   

\begin{figure}[h]
    \centering
    \includegraphics[width = 0.7\textwidth]{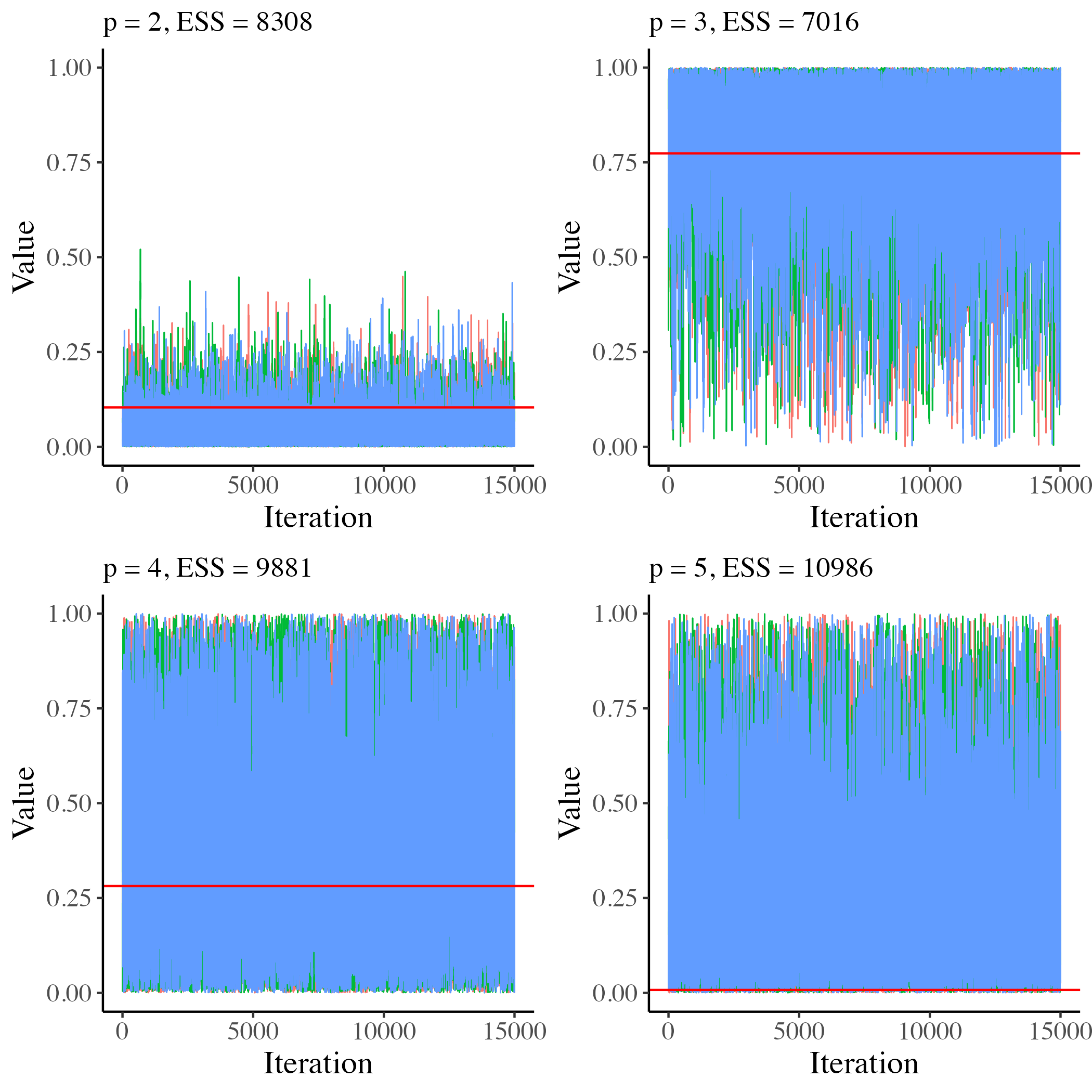}
    \caption{Method 1 (JAGS) Scenario 2 (The number of variables in $X$) - Trace plot of $\beta_1$}
    \label{fig:3:sat_p_tr}
    Trace plots and ESS of $\beta_1$. The red line denotes the true value of $\beta_1$. Among $5$ values of $p$, $p = 2,3,4$ and $5$ are plotted.
\end{figure}

In \autoref{fig:3:sat_p_tr}, the trend of ESS and trace plot differs from the other two scenarios. ESS decreases a bit and then increases again. Similarly, the trace plot doesn't show much autocorrelation from the smaller to larger $p$. However, even $p = 1$ takes $7.5$ minutes to complete, increasing to 100 minutes when $p = 5$. Therefore, even though the ESS and trace plot is reasonable, the ESS/time is almost $0$ in \autoref{fig:5:saturated_ess_time}. One noticeable thing in \autoref{fig:5:saturated_ess_time}: only multiESS shows as increasing from $p = 4$ to $p = 5$. However, this case's multiESS value is unreliable as the covariance matrix used to calculate it is close to singular. 

On the other hand, the method 2 performance worsens when $p$ increases, since ESS decreases from $30000$ to a small number around $0$ and time increases from $26$ seconds to $8$ minutes. However, ESS/time is still comparable to method 1, considering the poor performance of method 1 even in the smallest case of $p = 1$.

While scenario 1 and scenario 2 scale up the size of the problem by increasing the data size or parameter size, method 1 and method 2 show different trends in these two scenarios. Method 1 loses performance in terms of both ESS and time in scenario 1; however, only time is the issue in scenario 2. Method 2 doesn't lose performance in scenario 1 but worsens in scenario 2 both in ESS and time. However, this degradation is less than that of method 1.

\begin{table}[h]
\centering
\begin{tabular}{c|rrrrr}
  \hline
p & 1 & 2 & 3 & 4 & 5 \\   \hline
  Method 1 - JAGS & 459 & 784 & 1489 & 2901 & 5939 \\ 
  Method 2 - IM & 26 & 49 & 100 & 216 & 498 \\ 
   \hline
\end{tabular}
\caption{Simulation time of methods 1 and 2 (sec) as $p$ increases} 
\label{table:2:p_time} Method 1 uses JAGS and draws a sample from $\pi(\theta, y_{mis} \mid d_n)$. Method 2 uses ISTP and draws a sample from $\pi^*(\phi,\lambda \mid d_n)$.
\end{table}

\noindent\textbf{Scenario 3 ($\sigma = 0.01 - 10 ~ (7~ levels), n = 3000, p = 3$)}

\begin{figure}[h]
    \centering
    \includegraphics[width = 0.7\textwidth]{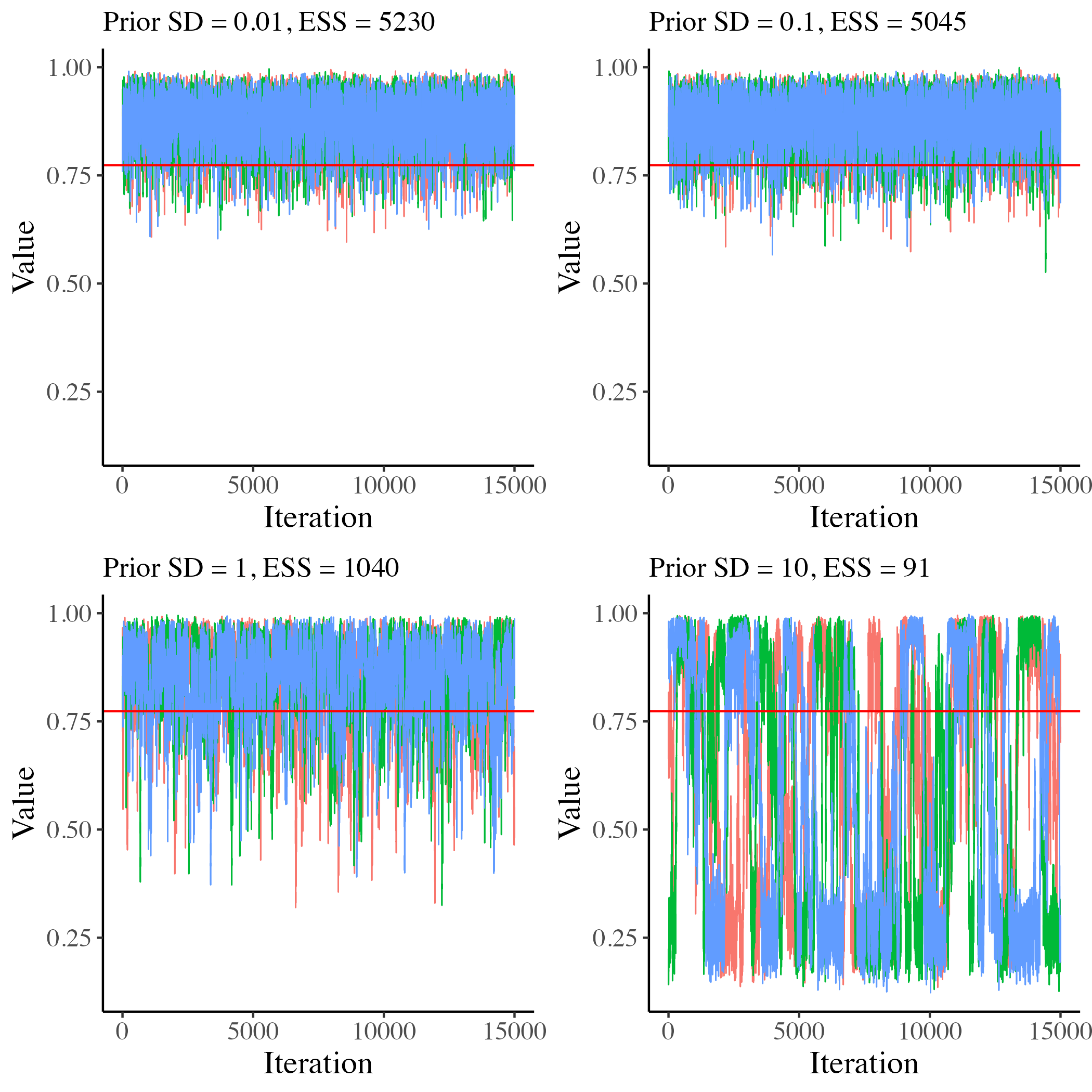}
    \caption{Method 1 (JAGS) Scenario 3 (Extent of MNAR) - Trace plot of $\beta_1$}
    \label{fig:4:sat_sigma_tr}
            Trace plots and ESS of $\beta_1$.  The red line denotes the true value of $\beta_1$. Among $7$ hyperparameters of $\sigma$, $\sigma = 0.01,0.1,1$ and $10$ are plotted.
\end{figure}

Scenario 3 is about the extent of MNAR allowed for by prior variances, which indicates how far the researcher feels the missing data mechanism could deviate from MAR. As the prior standard deviation, $\sigma$ increases, both method 1 and method 2 show worsening performance. In \autoref{table:3:sat_sigma_time}, time doesn't change much (around 23 minutes for method 1 and 1.7 minutes for method 2). Therefore, the worsening of performance is mainly in ESS. Both in \autoref{fig:4:sat_sigma_tr} and \autoref{fig:5:saturated_ess_time}, ESS decreases as prior SD increases. When comparing the results of methods 1 and 2, while the two methods show a trend of decreasing performance in larger $\sigma$, method 2 has much better performance because it has both a shorter time of sampling and larger ESS. The notable thing about scenario 3 is that unlike scenarios 1 and 2, method 1 shows a complete breakdown at the extreme case $\sigma = 10$. While method 2 also shows a steep decrease, it doesn't thoroughly crash as does method 1.

\begin{table}[h]
\centering
\begin{tabular}{c|rrrrrrr}
  \hline
$\sigma$ & 0.01 & 0.05 & 0.1 & 0.5 & 1 & 5 & 10  \\ 
  \hline
Method 1 & 1346 & 1359 & 1310 & 1325 & 1339 & 1389 & 1423 \\   Method 2 & 102 & 101 & 105 & 102 & 102 & 102  & 103 \\
   \hline
\end{tabular}
\caption{Simulation time of Method 1 and 2 (sec) as $\sigma$ increases} 
\label{table:3:sat_sigma_time}
Method 1 uses JAGS and draws a sample from $\pi(\theta, y_{mis} \mid d_n)$. Method 2 uses ISTP and draws a sample from $\pi^*(\phi,\lambda \mid d_n)$.
\end{table}


%% file: simplecount.tex
\newcommand{\xydes}{\mathbf{XY}_{des}}

\section{Count data with missing entries}\label{sec:4:count}
In the previous section, logistic regression applied to data with missing entries was explored. Since all variables were binary, finding a transparent reparameterization was not complicated. This process can be more challenging when the variables are not binary. In this section, we deal with count outcome data, which is discrete but not as simple as binary.
\begin{example}\label{ex:4:countreg} Count regression with incomplete outcome data: 
\begin{align*}
Y \mid \mathbf{X}_{des} & \sim Pois(\exp(\mathbf{X}_{des} {\beta})),\\
R \mid \mathbf{X}_{des}, Y = y
 &\sim Bern(\expit(\mathbf{X}_{des}(\gamma + \delta y))),
\end{align*}
\end{example}
Also, following the same definition as in other examples, $R$ is a binary indicator variable that is $1$ whenever $Y$ is available and is $0$ when $Y$ is missing.

The difference between \autoref{ex:3:logisticregsat} and \autoref{ex:4:countreg} is whether $Y$ is binary or count, and therefore, the reasonable assumption between $Y$ and $X$ is the linear regression in the transformed expectation of $Y$ given $X$. However, this makes a huge difference in finding an appropriate TP. This is because there is no guarantee that $Y$ given $R =1$ belongs to the same family of distributions (Poisson), as it does marginally. So, we introduce pseudo-TP. To test the computational performance with a pseudo-TP, we focus on the simpler example below:
\begin{example}\label{ex:5:simplecount}
    Estimating parameters in incomplete count data:
    \begin{equation}\label{eqn:9:ex5}
    \begin{aligned}
Y & \sim Pois(\mu),
\\
R \mid  Y & \sim Bern(\expit(\alpha_0 + \alpha_1 Y)).        
    \end{aligned}\end{equation}
\end{example}
\noindent The mock data are generated with fixed parameters of $\mu = 5, ~ \alpha_0 = -5\log(2)/\sqrt{5}  ~ \alpha_1 = \log(2)/\sqrt{5}$ by \texttt{R}. The parameter values $\alpha_0$ and $\alpha_1$ are chosen for standardizing as $E(Y) = Var(Y) = 5$. In particular, the constant with $Y$ being one SD below/above its mean induces on odds ratio of 2 for the chance of missingness. Bayesian analysis can be performed with uninformative priors:
\begin{align*}
\mu & \sim Gamma(1,1),\\
\alpha_0 & \sim Norm(0,10),\\
\alpha_1 & \sim Norm(0,\sigma^2).
\end{align*}
This \autoref{ex:5:simplecount} has no $X$, so it is much simpler now. Nontheless, it is not clear whether the model is fully identified or not. However, it is clear that this example still has nonignorable missingness \citep{little-2019-statistical} in $Y$, which may have a crucial impact on computational efficiency. At this point, a pseudo-TP might help, since data depends strongly on $\phi$ and weakly (or almost not) at all on $\lambda$.

\subsection{Pseudo-TP}
The structure of pseudo-TP is similar to the TP of \autoref{ex:3:logisticregsat}, but much simpler. The initial parameterization is $\theta = (\mu, \alpha_0, \alpha_1).$

In this example, the probability $p = Pr(R=1)$ and the conditional expectation $q = E(Y \mid R = 1)$ can be directly learned from the data. These two parameters are taken to comprise $\phi = (p,q)$.
Intuitively the data don't directly inform about $\alpha_1$ which describes the extent of MNAR. So, we can use it as $\lambda$ to represent the impact of the extent of MNAR.
As discussed in \autoref{ex:4:countreg}, the closed-form TP may not exist. Even if we could define $h(\theta) = (\phi,\lambda)$ where $\phi = (p,q)$ and $\lambda = \alpha_1$, there is no closed-form function for $h^{-1}(\phi,\lambda)$.

But with the assumption of $(Y \mid R=1)$ following Poisson distribution and Taylor approximations for finding some necessary parameters, we may find a reparameterization close to TP.
Now we find $h_*(\theta)$ and $h^{-1}_*(\phi,\lambda)$ instead, functions for pseudo-TP. 
\noindent $h_*(\theta)$ is based on the following approximations:
\allowdisplaybreaks
\begin{align*}
p = Pr(R = 1) &= \sum_{y = 0}^{\infty} Pr(Y = y) Pr(R = 1 \mid Y = y)\\
					 &=E(Pr(R = 1 \mid Y))\\
					 & = E(\expit(\alpha_0 + \alpha_1 Y))\\
					 & \approx \expit(\alpha_0 + \alpha_1 E(Y))\\
					 & = \expit(\alpha_0 + \alpha_1 \mu) \\&:= p^*\\\\
q  = E(Y \mid R = 1) &= \frac{\sum_{y=0}^\infty y \cdot Pr(Y = y , R = 1)}{Pr(R=1)} \\
&= \frac{\sum_{y=0}^\infty y \cdot Pr(Y = y , R = 1)}{\sum_{y = 0}^{\infty} Pr(Y = y) \cdot Pr(R = 1 \mid Y = y)}\\
&= \frac{\sum_{y=0}^\infty y \cdot Pr(R = 1 \mid Y = y) \cdot Pr(Y = y)}{\sum_{y = 0}^{\infty} Pr(Y = y) \cdot Pr(R = 1 \mid Y = y)}\\
&= \frac{\sum_{y=0}^\infty y \cdot \expit (\alpha_0 + \alpha_1 y) \cdot \frac{e^{-\mu} \mu^y}{y!} }{\sum_{y = 0}^{\infty} \expit (\alpha_0 + \alpha_1 y) \cdot \frac{e^{-\mu} \mu^y}{y!}}\\
&= \frac{\sum_{y=0}^\infty y \cdot \exp(g(y)) \cdot \frac{\exp(-\mu) \mu^y}{y!} }{\sum_{y = 0}^{\infty} \exp(g(y))\cdot \frac{\exp(-\mu) \mu^y}{y!}}\\
&\approx \frac{\sum_{y=0}^\infty y \cdot \exp({g(\mu) + g^{'}(\mu)(y-\mu)})\cdot \frac{\exp(-\mu)  \mu^y}{y!} }{\sum_{y = 0}^{\infty} \exp({g(\mu) + g^{'}(\mu)(y-\mu)})\cdot \frac{\exp(-\mu)  \mu^y}{y!}}\\
&= \frac{\sum_{y=0}^\infty y \cdot \exp({yg^{'}(\mu)}) \cdot \frac{ \mu^y}{y!} }{\sum_{y = 0}^{\infty} \exp({yg^{'}(\mu)})\cdot \frac{ \mu^y}{y!}}\\
&= \frac{\sum_{y=0}^\infty y  \cdot \frac{ (\mu \exp({g^{'}(\mu)}))^y}{y!} }{\sum_{y = 0}^{\infty} \frac{(\mu \exp({g^{'}(\mu)}))^y}{y!}}\\
&= \frac{\sum_{y=0}^\infty y  \cdot \frac{ (\mu^{*})^y}{y!} }{\sum_{y = 0}^{\infty} \frac{(\mu^{*})^y}{y!}}\\
&= \frac{\sum_{y=0}^\infty y  \cdot \frac{ \exp(-\mu^*)(\mu^{*})^y}{y!} }{\sum_{y = 0}^{\infty} \frac{ \exp(-\mu^*)(\mu^{*})^y}{y!}}\\ 
&= \frac{\mu^*}{1}\\
&= {\mu^*} \\ &:=q^* .
\end{align*}
where $g(y) = \log (\expit(\alpha_0 + \alpha_1 y))$ and $\mu^* = \mu \exp({g^{'}(\mu)}) = \mu (\exp(\alpha_1\cdot(1-\expit(\alpha_0 + \alpha_1\mu)))$. By the approximation of $p$, which is $p^*$, above, $q^* = \mu^* \approx \mu (\exp(\alpha_1\cdot(1-p^*)))$. While we learn $ (p,q)$ from the data, approximations $p^*$ and $q^*$ are used for finding a closed-form inverse function.

Since Taylor approximation is used multiple times, $h_*(\theta)$ is not a TP, but we expect that it is close to a TP since $h_*(\theta)$ is a legitimate approximation. Then, the inverse function $h^{-1}_*(\phi,\lambda)$ is given by the system of equations  below:
\begin{align*}
  \mu & = \mu^* / (\exp(\alpha_1\cdot(1-p^*)))    \\
  \alpha_0 & = \logit(p^*) - \alpha_1 \cdot \mu  
\end{align*}

\subsection{Simulation methods and results}
In this section, for testing the performance of sampling with pseudo-TP, we apply simulations to two scenarios introduced in \autoref{sec:3.2.:satsim}, one with increasing the number of observations and the other with further deviation from MAR. The scenario with increasing $p$ is excluded since this example is inherently univariate. Scenario 1 uses data with $n =3$ to $n =10000$ while fixing $\sigma = 0.5$. And in scenario 2, a narrow prior ($\sigma = 0.01$) to a wide prior ($\sigma = 10$) are used, where $\sigma^2$ is the prior variance of $\alpha_1$. For all simulations in scenario 2, $n = 3000$.

While method 1 remains the same as \autoref{sec:3.2.:satsim} with traditional MCMC using JAGS applied in the original parameterization, two methods are used for testing pseudo-TP. 
\begin{itemize}
    \item Method 2

    This method uses JAGS. But convenience priors are specified for $\phi$ and $\lambda$. 
\begin{align*}
    p & \sim Beta(1,1)\\
q & \sim Gamma(1,1)\\
\alpha_1 & \sim Norm(0,\sigma^2)
\end{align*}
    
    Then, $\theta$ is found with $\theta = h_*^{-1}(\phi,\lambda)$ and JAGS updates the states of $\theta$ with the true likelihood as in \autoref{ex:3:logisticregsat}.

    The main difference between methods 1 versus 2 is that JAGS iterates on $\theta$ (method 1) and $(\phi,\lambda)$ with mapping back to $\theta$ (method 2). Even though method 2 couldn't thoroughly reparameterize $\theta$, this difference might alleviate the computation efficiency issue arising from partial or weak identification.

    \item Method 3

    This method uses importance sampling with pseudo-TP. The crucial point of this method is facilitating i.i.d. sampling. This is achieved by 
using a convenience likelihood as well as a convenience prior. The convenience likelihood is constructed under the assumption that $(Y \mid R = 1)$ follow the Poisson distribution, while method 2 is free from the distributional assumption. Consequently, the parts of likelihood in weight ratio can't be cancelled since the likelihood in with the assumption is no longer equivalent to the original likelihood. Consequently, the weight ratio is found with 
    $W_i = \frac{\pi(\phi_i, \lambda_i) l (d_n \mid \phi_i, \lambda_i)}{\pi^*(\lambda_i \mid \phi_i) \cdot \pi^* (\phi_i) \cdot l^* (d_n \mid \phi_i)}$ where $l^* (d_n \mid \phi_i)$ is convenience likelihood and $l (d_n \mid \phi_i, \lambda_i)$ is true likelihood. The convenience likelihood is \begin{equation}
        \begin{aligned}
            R & \sim Bern(p),\\
            (Y &\mid R = 1) \sim Pois(q).
        \end{aligned}
    \end{equation}   
    \noindent where $p$ and $q$ are as in the previous definition of parameters. The importance weight of ordinary TP doesn't have likelihood parts and would not require data. However, since the true likelihood $l(d_n\mid\theta)$ in \autoref{eqn:9:ex5} requires all entries of $y$, imputing of $Y$ is necessary. JAGS in methods 1 and 2 imputes missing entries by default, but in method 3, we first find the conditional expectation $E(Y \mid R =0) \approx \mu \cdot \exp(-\alpha_1 \cdot \expit(\alpha_0 + \alpha_1 \mu))$ and impute $Y$ on the assumption that $Y \mid R = 0$ follows the corresponding Poisson distribution.
\end{itemize}
\large{\noindent\textbf{Results}}

\noindent Following the same process as in \autoref{sec:3.2.:satsim}, we look into ESS, simulation time, and trace plot of target parameter, $\mu = E(Y)$.

\noindent\textbf{Scenario 1}

\begin{figure}[h]
    \centering
    \includegraphics[width = 0.88\textwidth]{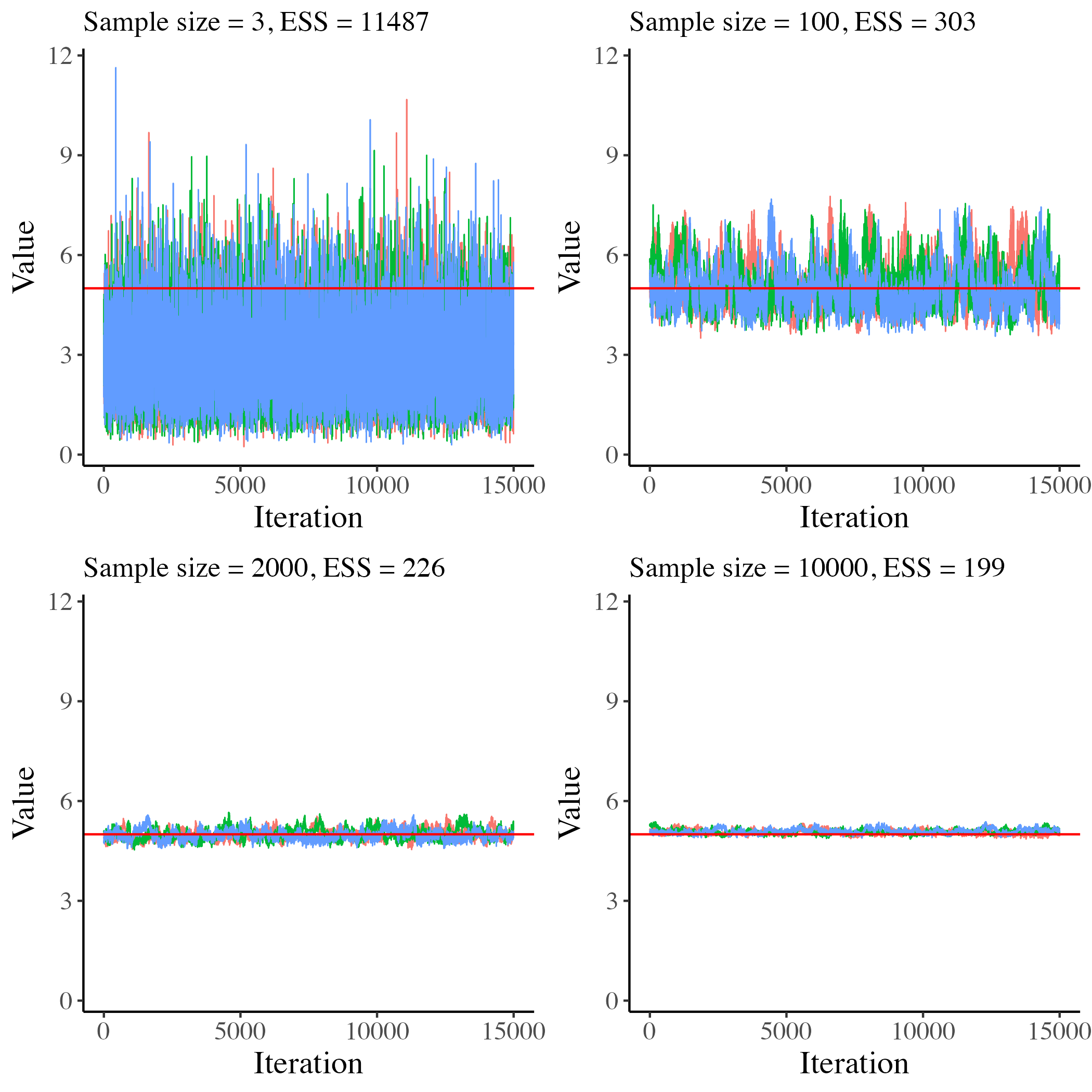}
    \caption{Method 1 (JAGS, original) Scenario 1 (The number of observations) - Trace plot}
    \label{fig:6:jags_o_n_tr}
\end{figure}

\begin{figure}
    \centering
    \includegraphics[width = 0.9\textwidth]{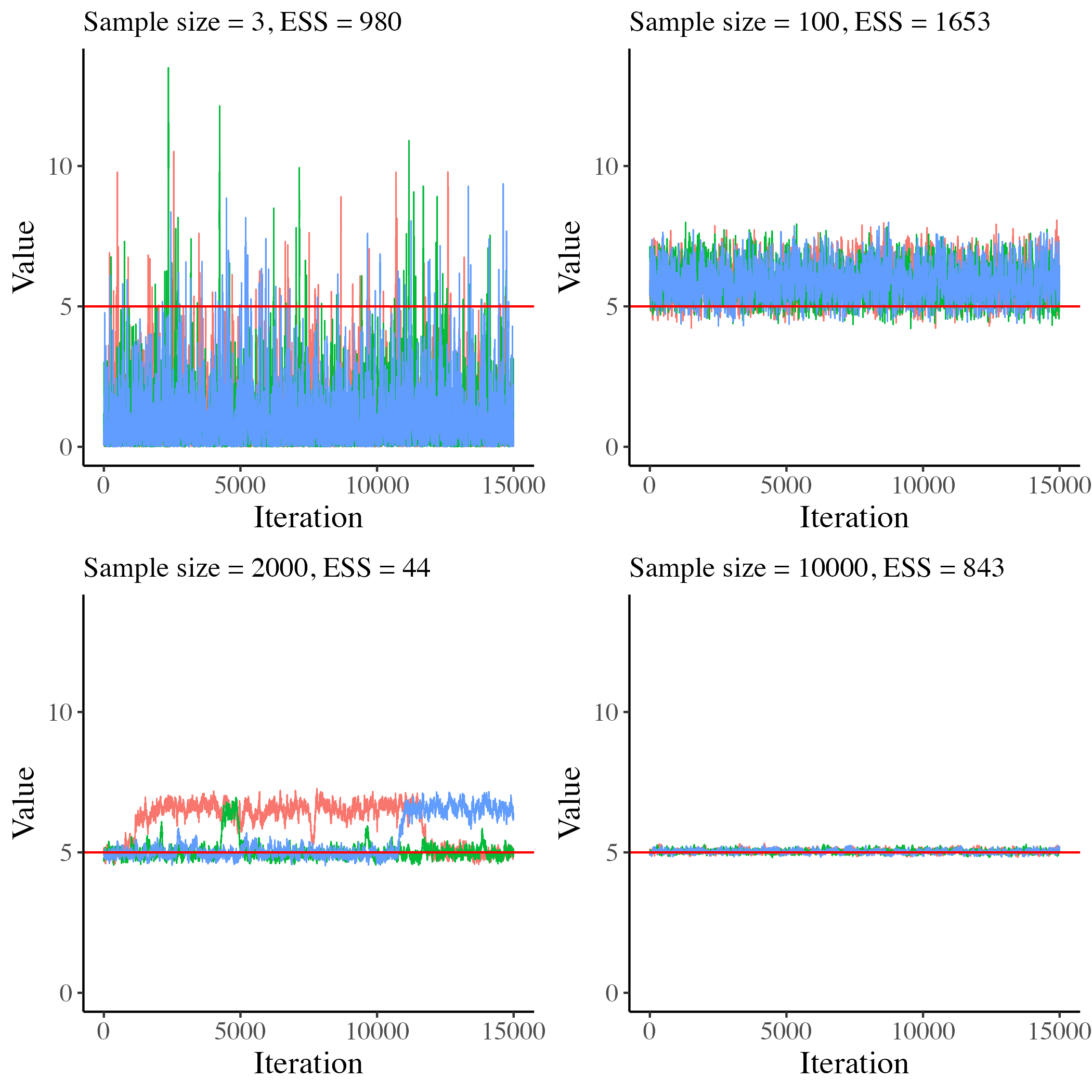}
    \caption{Method 2 (JAGS, Pseudo-TP) Scenario 1 (The number of observations) - Trace plot}
    \label{fig:7:jags_r_n_tr}
\end{figure}

\begin{figure}
    \centering
    \includegraphics[width = 0.49\textwidth]{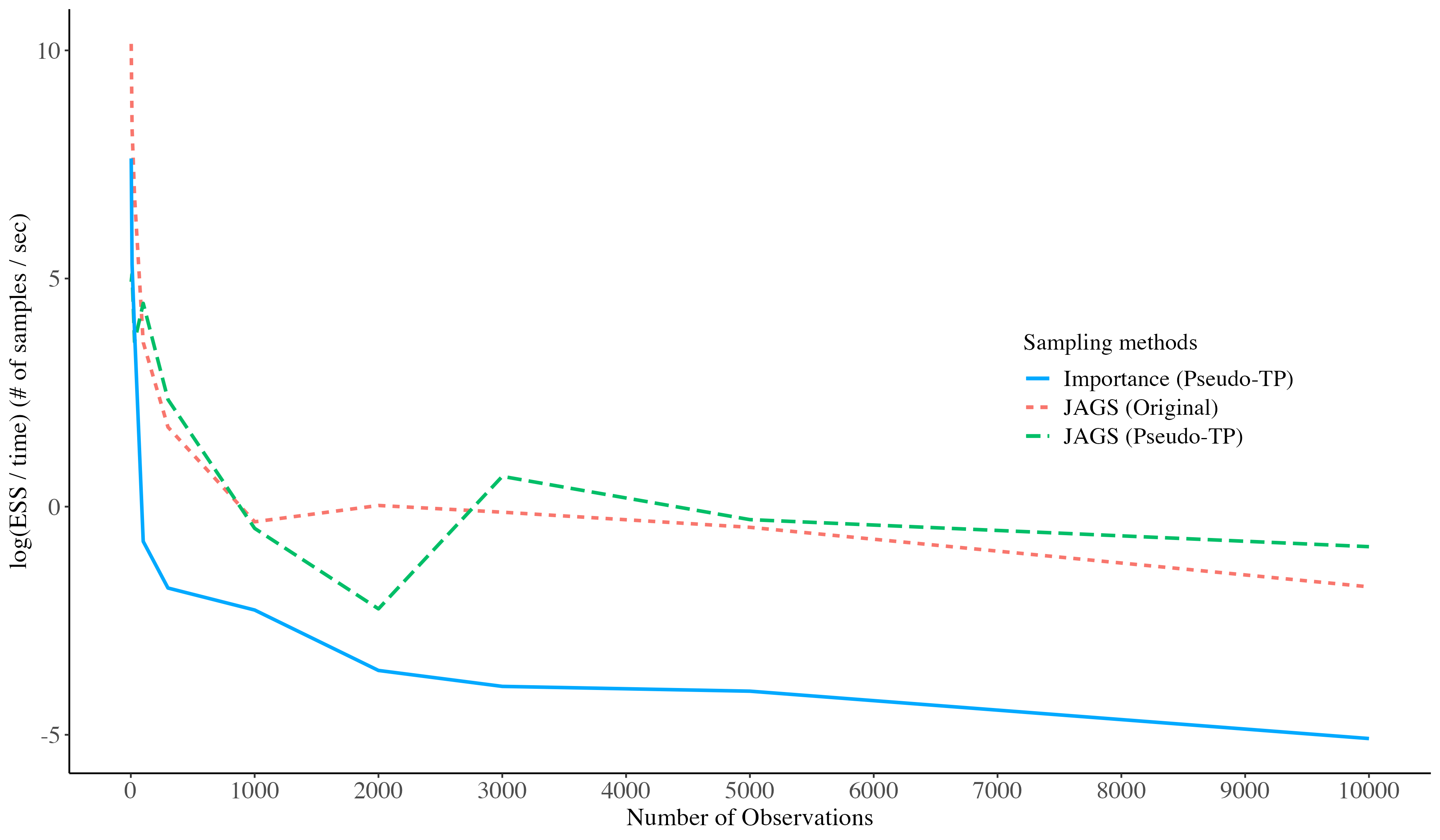}
    \includegraphics[width = 0.49\textwidth]{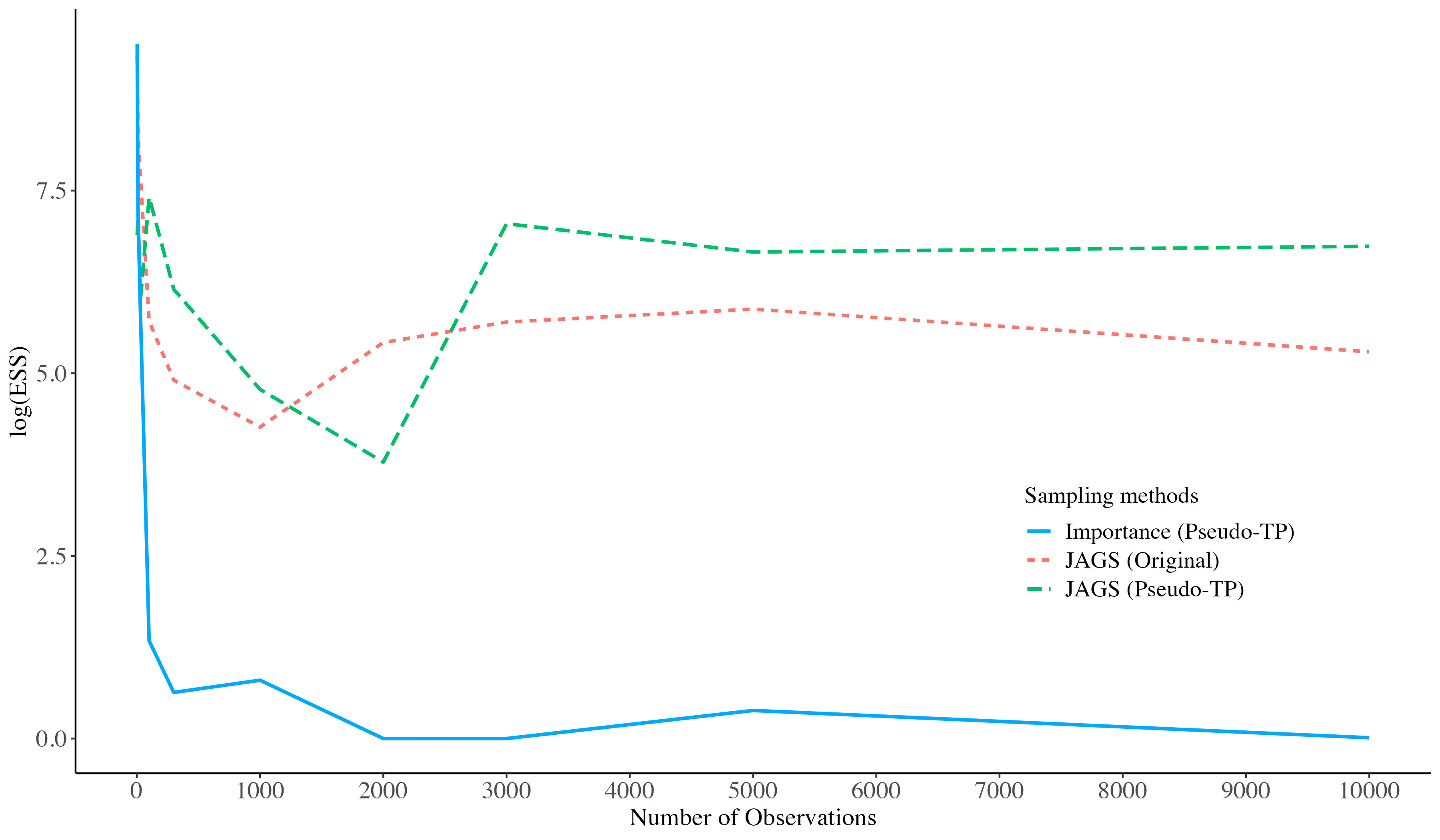}
    \includegraphics[width = 0.49\textwidth]{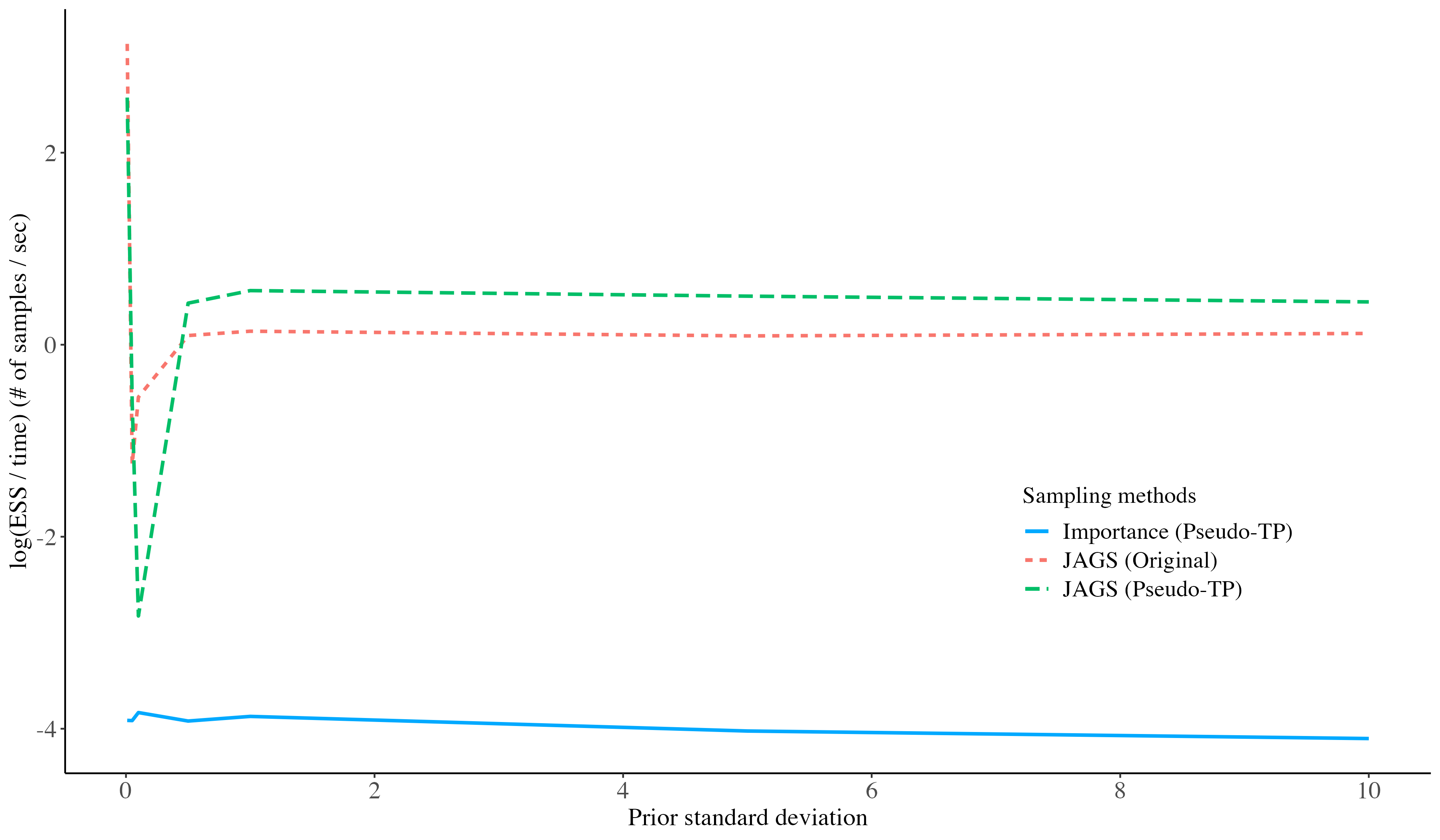}
    \includegraphics[width = 0.49\textwidth]{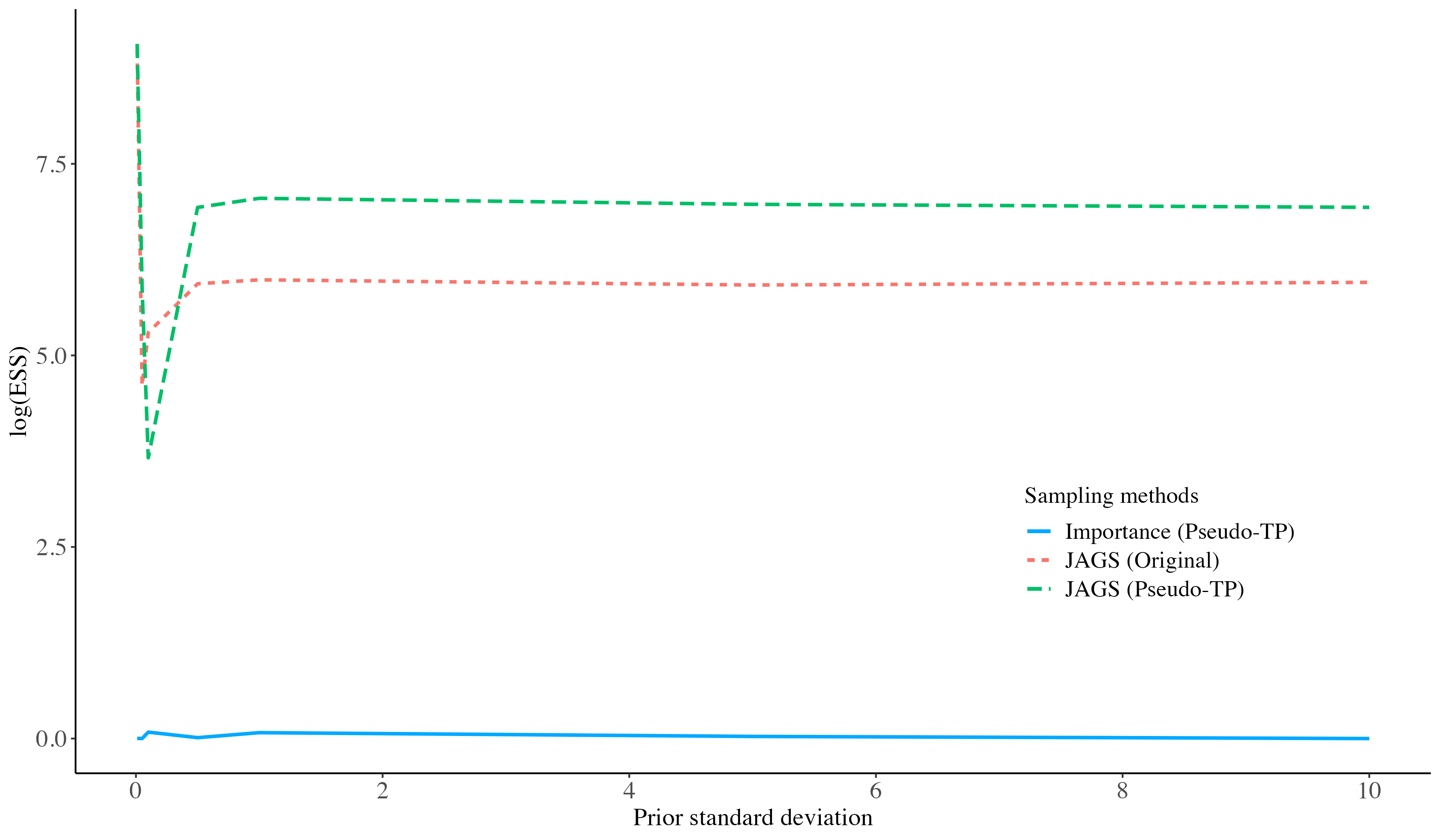}
    \caption{ESS and time in each method and scenario}
    \label{fig:10:count_simple_ess_time}
\end{figure}

In \autoref{fig:6:jags_o_n_tr}, we can see that ESS of $\mu$ first steeply decreases from 11487 to 303 and slightly decreases, as $n$ increases from 3 to 10000, when using method 1. Also, as time increases rapidly, ESS/time converges to $0$ as $n$ increases. Method 2 results have larger ESS when $n$ is larger than 2000 in \autoref{fig:10:count_simple_ess_time}; however, method 2 doesn't have predominance in ESS/time due to longer performance time as in \autoref{table:4:simple_n_time}. 
One noticeable thing is that in both \autoref{fig:6:jags_o_n_tr} and \autoref{fig:7:jags_r_n_tr}, the bottom-right plot $(n = 10000)$ shows strong evidence for the hypothesis that the model is identified since the posterior variance of target parameter $\mu$ is very small.
It is not clear whether the model is mathematically identified or not. And the plots are suggestive of it being identified, albeit weakly identified.

On the other hand, method 3 shows the lowest ESS/time and ESS among the three methods. Even though method 3 only took a few seconds to a few minutes to execute the algorithm, ESS is usually near 1. Its behaviour is different from the results in \autoref{ex:3:logisticregsat}, where ISTP has better performance both in terms of ESS and time.

\begin{table}[ht]
\centering
\begin{tabular}{c|rrrrrrrrrr}
  \hline
 $n$ & 3 & 10 & 30 & 100 & 300 & 1000 & 2000 & 3000 & 5000 & 10000  \\ 
  \hline 
 Method 1 & 0.5 & 1.0 & 2.4 & 8.3 & 23.6 & 98.9 & 220.3 & 337.5 & 561.4 & 1153.9 \\   Method 2 & 7.1 & 8.0 & 11.1 & 19.2 & 44.7 & 191.5 & 413.3 & 591.5 & 1036.4 & 2026.6 \\   Method 3 & 6.5 & 6.7 & 7.1 & 8.2 & 11.2 & 21.4 & 36.3 & 51.4 & 83.8 & 163.0 \\
   \hline
\end{tabular}
\caption{Simulation time of methods 1 - 3 (sec) as $n$ increases} 
\label{table:4:simple_n_time}
\end{table}

\noindent\textbf{Scenario 2}

\begin{figure}
    \centering
    \includegraphics[width = 0.9\textwidth]{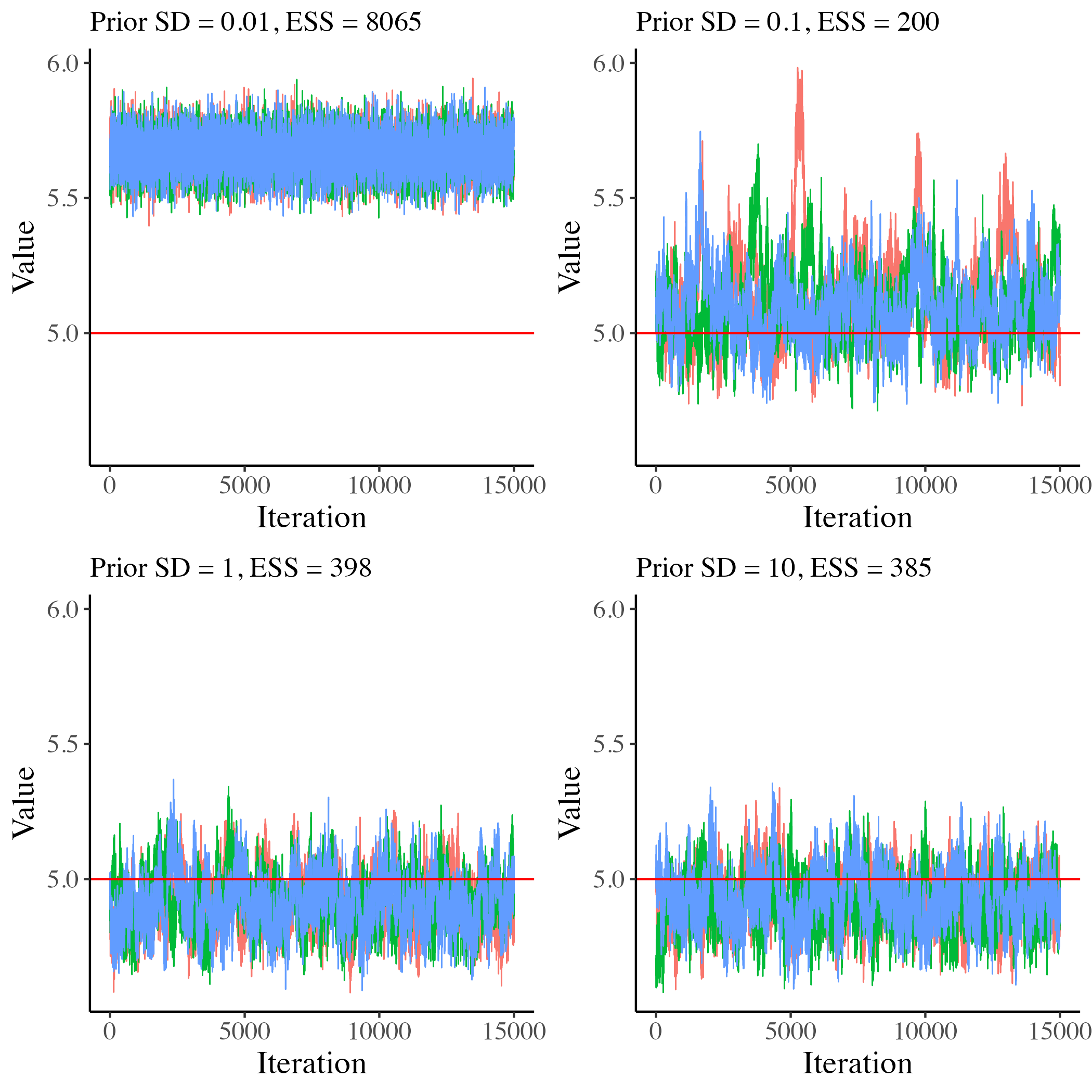}
    \caption{Method 1 (JAGS, original) Scenario 2 (Extent of MNAR) - Trace plot}
    \label{fig:8:jags_o_sigma_tr}
\end{figure}

\begin{figure}
    \centering
    \includegraphics[width = 0.9\textwidth]{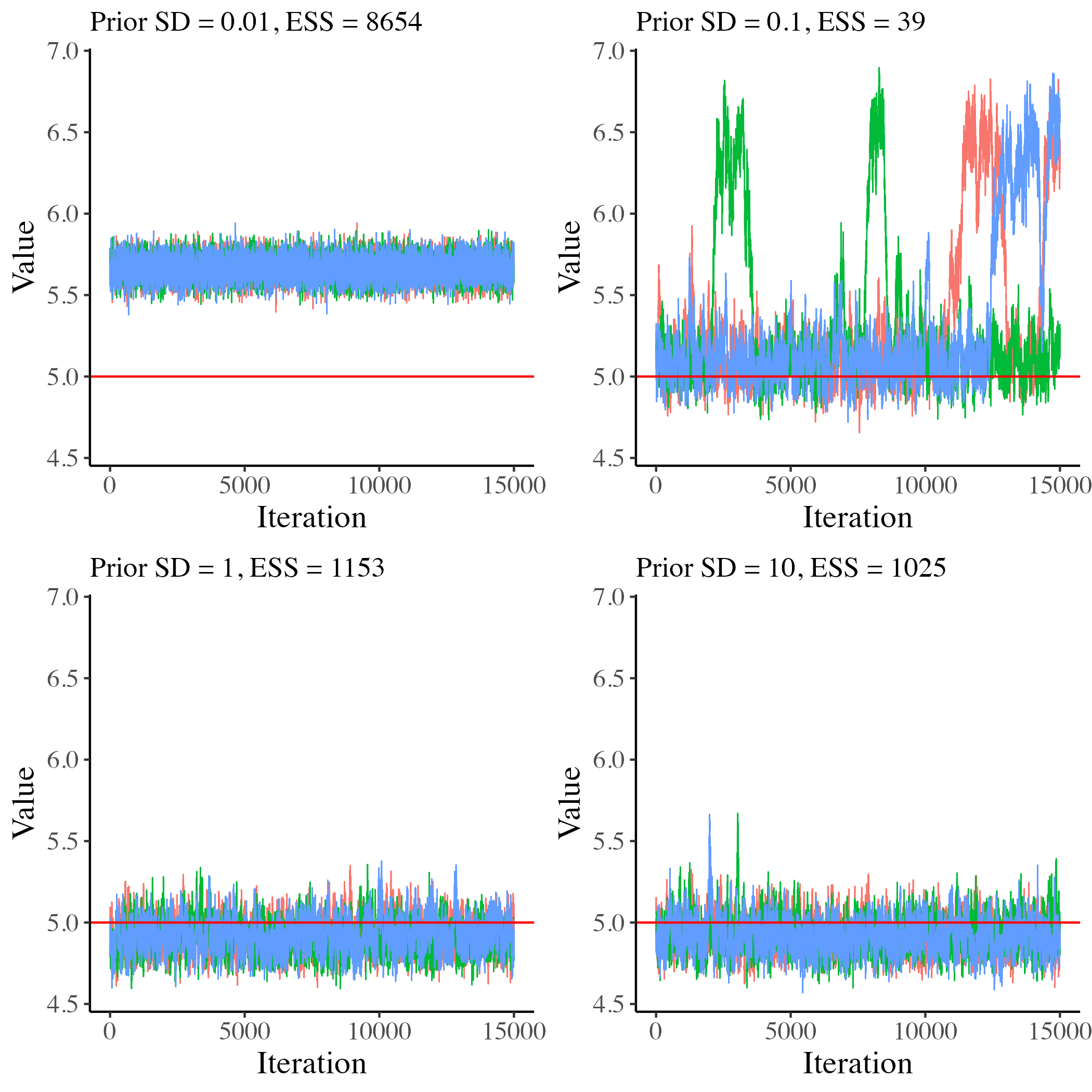}
    \caption{Method 2 (JAGS, Pseudo-TP) Scenario 2 (Extent of MNAR) - Trace plot}
    \label{fig:9:jags_r_sigma_tr}
\end{figure}

The results are a bit different in scenario 2. We expect the performance will worsen prior when the distribution allows further deviation from MAR as in \autoref{ex:3:logisticregsat}. However, from $\sigma = 0.01$ to $\sigma = 0.1$, ESS first decreases steeply, and the trace plot shows autocorrelation or chains are not overlapped in \autoref{fig:8:jags_o_sigma_tr} and \autoref{fig:9:jags_r_sigma_tr}. Then, ESS remains similar after a small increase in both methods 1 and 2. Method 2 generally shows larger ESS and larger ESS/time, but ESS/time results are not much larger in method 2 due to the longer simulation time as seen in \autoref{fig:10:count_simple_ess_time} and \autoref{table:5:simple_sigma_time}. Again, method 3 has a small ESS and ESS/time.
\begin{table}[ht]
\centering
\begin{tabular}{c|rrrrrrr}
  \hline
 $\sigma$ & 0.01 & 0.05 & 0.1 & 0.5 & 1 & 5 & 10  \\ 
  \hline
Method 1 & 351.3 & 349.3 & 345.3 & 343.5 & 345.8 & 339.4 & 342.4 \\ Method 2 & 659.4 & 654.4 & 658.0 & 664.2 & 656.2 & 642.8 & 656.2 \\ Method 3 & 50.2 & 50.2 & 50.2 & 51.0 & 51.8 & 57.5 & 60.5 \\
   \hline
\end{tabular}
\caption{Simulation time of methods 1 - 3 (sec) as $\sigma$ increases} 
\label{table:5:simple_sigma_time}
\end{table}

We suspect that the reasons behind the poor performance of method 3 are:
\begin{itemize}
    \item Reparameterization is based on an approximation using the first-order Taylor expansion. 
    \item Compared to method 2, there is a further approximation that $(Y \mid R = 1)$ follows the Poisson distribution. The weight in method 3 takes this approximation into account by not cancelling the ratio of true to convenience likelihood. Our results show that the convenience likelihood approximation is not close enough to the true likelihood for effective importance weighting.  To find the weight, ratios between two posterior densities are used. When finding them, an imputation of $Y$ is applied to get the likelihoods. This imputation has two approximations/assumptions. $E(Y \mid R = 0) = \mu \cdot exp(-\alpha_1 \cdot \expit(\alpha_0 + \alpha_1 \mu))$ is found with the first order Taylor expansion, and $(Y \mid R = 0)$ is assumed to follow the Poisson distribution.
    Most normalized weights are near 0, and only a few or one is almost 1. The imputed data convenience likelihood and true likelihood are both for the joint distribution of $(Y, R)$, and this problem arises from the overall discrepancy between two joint-likelihoods $f(y,r)$ and $f^*(y,r)$.
 \end{itemize}

%% file: concldiss.tex
\section{Discussion}
In this paper, we tested transparent reparameterization and pseudo-TP in some scenarios of a model either known or suspected to be partially identified. Since partially identified models can be easily set in ways that scientifically make sense, it would be helpful if TP and pseudo-TP could resolve the issue of poor sampling performance with traditional MCMC. When a TP is available, it is shown that there is an improvement in the computational performance by using ISTP, especially in terms of compute time. However, when TP is unavailable, and consequently, pseudo-TP is used instead, IM does not work. The issue arises from a gap between actual and convenience likelihood. In this section, we discuss the performance of importance sampling with TP/pseudo-TP, its challenges, improvements, and future topics.

\subsection{Performance of sampling with TP and Pseudo-TP in examples}
\subsubsection{Example 3 - Transparent reparameterization in saturated logistic regression}
In \autoref{sec:3:sat}, we tested some scenarios involving a partially identified model. When an off-the-shelf MCMC algorithm is applied to draw a sample from the posterior distribution in the partially identified model, it is expected to break down. Two scenarios focus on the size of the problem. Since the MCMC method, especially Gibbs sampling (JAGS, method 1), updates the chain of parameters based on data, the computational performance is expected to worsen when the size of the problem is larger. Indeed, when the number of observations in the data, $n$, or the number of parameters in $X$, $p$, increase, sampling time increases steeply. Also, in scenario 1, where $n$ is varied, ESS is small in not trivially small $n$ for method 1. The large ESS of method 1 in small $n$ is mainly because the posteriors are close to priors, which don't have much autocorrelation.

On the other hand, as ISTP (method 2) allows the usage of i.i.d. Monte Carlo sampling in \autoref{ex:3:logisticregsat}, the number of observations is not the issue for computational performance.  Similarly, increasing the number of parameters $p$ doesn't affect much but shows that ESS is almost 0 when $p =5$, where the dimension of $\theta$ is over $100$. Regardless of ESS, the simulation time is generally stable around a few seconds to a minute in all scenarios. While the number of parameters in $X$, $p$, impacts the weight calculation time, this will not be a big problem unless the number of total parameters in the problem is unrealistically large.

The other scenario concerns the prior variance, $\sigma^2$, representing the deviation from MAR to MNAR. We found that both ESS/time and ESS decrease as $\sigma$ increases. While the results of method 2 also show the break down of ESS in the large variance case of $\sigma^2 = 100$, ESS/time is larger than method 1 for all variance values.

There are two main concerns: i) ESS and ii) Compute time. Generally, method 2 shows better performance than method 1 in terms of these concerns. This somewhat agrees with the comments about ISTP in \citep{Pirikahu-2021-Bayesian}. When the authors explained this method, they mentioned that ISTP could be an effective alternative for MCMC methods in problems with a partially identified model. However, they were also concerned that ISTP relies on the availability of finding a TP and finding this TP may not be possible or easy, especially when there are many parameters in the problem. In this problem, even though method 2 is not working well when there is a large number of parameters, it is not about finding a reparameterization but instead about the dimensions of parameters. Since parameters are structured, finding the TP when $p$ is large is not hard. It would still be the same for extremely large $p$ as long as parameters are structured. However, as the authors commented, ISTP can only be implemented when researchers can find an appropriate TP.

\subsubsection{Example 5 - Pseudo-TP in the count data}
The best scenario is for a TP to be available for the given problem. However, as seen in \autoref{sec:4:count}, even a small change in the problem can make it hard or impossible to find a TP.
However, if we find a reparameterization that behaves similarly to a TP, it is still worth considering whether it improves computational performance, while acknowledging the drawback arising from not being a TP.
In \autoref{ex:5:simplecount}, we found pseudo-TP based on Taylor approximation. Then, we performed simulations in three ways: Method 1: JAGS with original parameterization, Method 2: JAGS with pseudo-TP, using a convenience prior and the true likelihood $l(d_n\mid\theta)$, and Method 3: IM with pseudo-TP, using a convenience prior and a convenience likelihood. 
Those functions are found using the first-order Taylor approximation. Method 2 performs better than method 1 in terms of ESS and trace plots. However, the compute time is almost doubled, and still, the true likelihood $l(d_n\mid \theta)$ used for updating the states of $\theta$ has the autocorrelation issue in parameters that explain $R$ and $X, Y$. In the true likelihood $l(d_n\mid \theta)$, $R$ is still explaining the existence of $Y$ while $Y$ is explaining the probability of $R$ being 1. So, we expected that IM with pseudo-TP can be a good option since method 3 will use i.i.d. Monte Carlo sampling using \texttt{R} base functions is faster than MCMC sampling. However, to use the i.i.d. Monte Carlo sampling, another assumption is required for the conjugate convenience priors. While $q = E(Y \mid R = 1)$ is found with the Taylor approximation, it does not give any information regarding the distribution of $Y \mid R = 1$. It is the same for $Y \mid R = 0$, which is required for evaluating the true likelihood $l(d_n \mid \theta)$. Therefore, simulation method 3 is based on the assumption of $(Y \mid R = 1)$ and $(Y \mid R = 0)$ following the Poisson distribution. With these assumptions, IM with pseudo-TP fails, as ESS is too small. As briefly discussed in the previous section, the main reason is that the gap between true and convenience likelihood is too large. Method 2 and 3 have the trade-off between the probably more accurate computational result by using less approximation and the saving of time by using conjugate convenience prior, which enables i.i.d. Monte Carlo sampling. At least in this example, unless the issue of tiny ESS is resolved by reducing the gap between true and convenience likelihood, even if method 2 has kind of a mixed result - better ESS and trace plot but longer simulation time, method 3 can't be an alternative for method 2, or even method 1.

The other notable thing in \autoref{ex:5:simplecount} is the challenge in discriminating between a partially identified versus a weakly identified model. In \autoref{sec:2.1:pim}, we discussed that mathematically, it is hard to distinguish those two types of models. It is not clear whether this model for count data is fully identified or partially identified. After performing simulations, we found that as
the number of observations increases, the variance of a posterior sample becomes really small. This happens when $\alpha_1$ is learned through data, which suggests identification of the model. However, at the same time, the ESS decreases significantly compared to the result of small $n$, which is the expected behaviour of a partially identified model.

\subsection{Further research topics}
\subsubsection{Diagnosis tool for the existence of TP}
There is no algorithmic decision rule on whether a TP exists for a given MCMC problem. 
Even if the investigator can't find a TP, that doesn't necessarily mean that a TP doesn't exist. 
From the researchers' point of view, it is essential to decide whether to proceed further to find a TP in terms of efficiency in the computational process. 
For that purpose, a diagnostic tool for checking the existence of TP could play a crucial role. 
In \autoref{sec:2.2.1:tp}, there are two conditions that a transformation needs to meet to be transparent reparameterization. However, these conditions can only be checked after a putative transformation is found. 
So, these conditions are necessary conditions for some transformation to be a TP, not necessary conditions for a given problem to have a TP. 
If we can find the necessary conditions for the given problem to have TP, it will be easier to check whether TP is available without ad-hoc attempts to find a TP.  We surmise that these necessary conditions are related to the Jacobian matrix and Fisher information of the parameters in the problem. 

\subsubsection{Improved version of pseudo-TP}
The concept of a pseudo-TP has a lot of room for improvement. Since pseudo-TP is a broad term that covers any transformation that is not a TP but has some intuitive flavour of a TP, a given pseudo-TP may work well, or may fail to work because it relies on overly rough approximations. Therefore, the goal is to minimize the gap arising from approximation and assumption while still ensuring the reparameterizing transformation is invertible. 

One possible way forward is to use an incremental or adaptive method. In \autoref{ex:5:simplecount}, at least two approximations are combined: a Taylor approximation and a Poisson distributional assumption. The impact of all assumptions is not the simple summation of the effects of each assumption. As seen in the simulation results of method 2, at least the Taylor approximation doesn't break the performance of posterior sampling. However, simulation results show that method 3 fails to work. The difference between methods 2 and 3 is the addition of the Poisson assumption. It is appealing to use conjugate convenience prior and, therefore, use i.i.d. Monte Carlo sampling in IM. However, if the IM with pseudo-TP can apply approximations one by one gradually instead of applying two approximations simultaneously, the sampling performance may get better. 
\subsubsection{Weakly vs Partially identified model}
In \autoref{ex:5:simplecount}, the model is not determined to be fully identified or partially identified. We patently recognize the date has the issue of nonignorable missingness, but that doesn't necessarily indicate that the model is partially identified. Indeed, the model seems weakly identified in \autoref{fig:6:jags_o_n_tr} when the sample size is large. However, this information is available only after the actual MCMC sampling is done and is not easily recognizable in the model specification stage. 

If we use the information that can be learned directly from the data, we might prove the identification of the model. In this \autoref{ex:5:simplecount}, it is possible to learn the probability $c_y = Pr(Y =y, R = 1) = Pr(R = 1 \mid Y = y) Pr(Y=y)$ for any non-negative integer $y$. And, we can easily find one-to-one function between $(a_0,a_1)$ and $(\alpha_0,\alpha_1)$ where $a_i = Pr(R = 1 \mid Y= y) = \expit(\alpha_0 + \alpha_1)$. Then, it is possible to represent the probability $Pr(R = 1 \mid Y = 2)$ as a function of $(a_0,a_1)$. Let this function be $h(a_0,a_1) = \expit(2\logit(a_1) - \logit(a_0)).$ The identification of the model is implicated by the number of solutions $(\mu, a_0, a_1)$ in the system of equations:
\begin{align*}
    c_0 &= e^{-\mu} a_0 \\
    c_1 &= e^{-\mu} \mu a_1 \\
    c_2 &= \frac{e^{-\mu}\mu^2}{2}  h(a_0,a_1).
\end{align*}
\begin{figure}
    \centering
\includegraphics[width=0.4\linewidth]{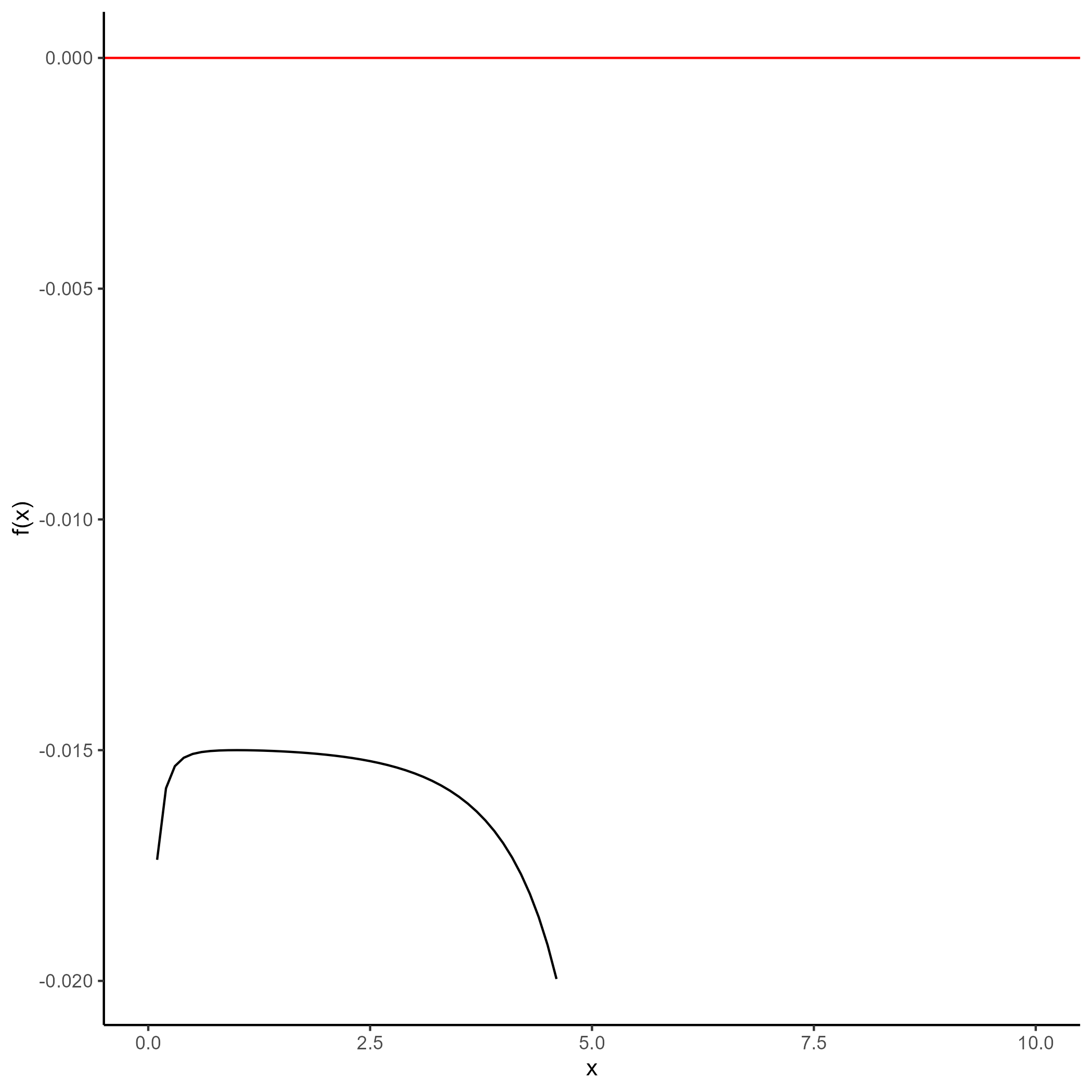}\includegraphics[width=0.4\linewidth]{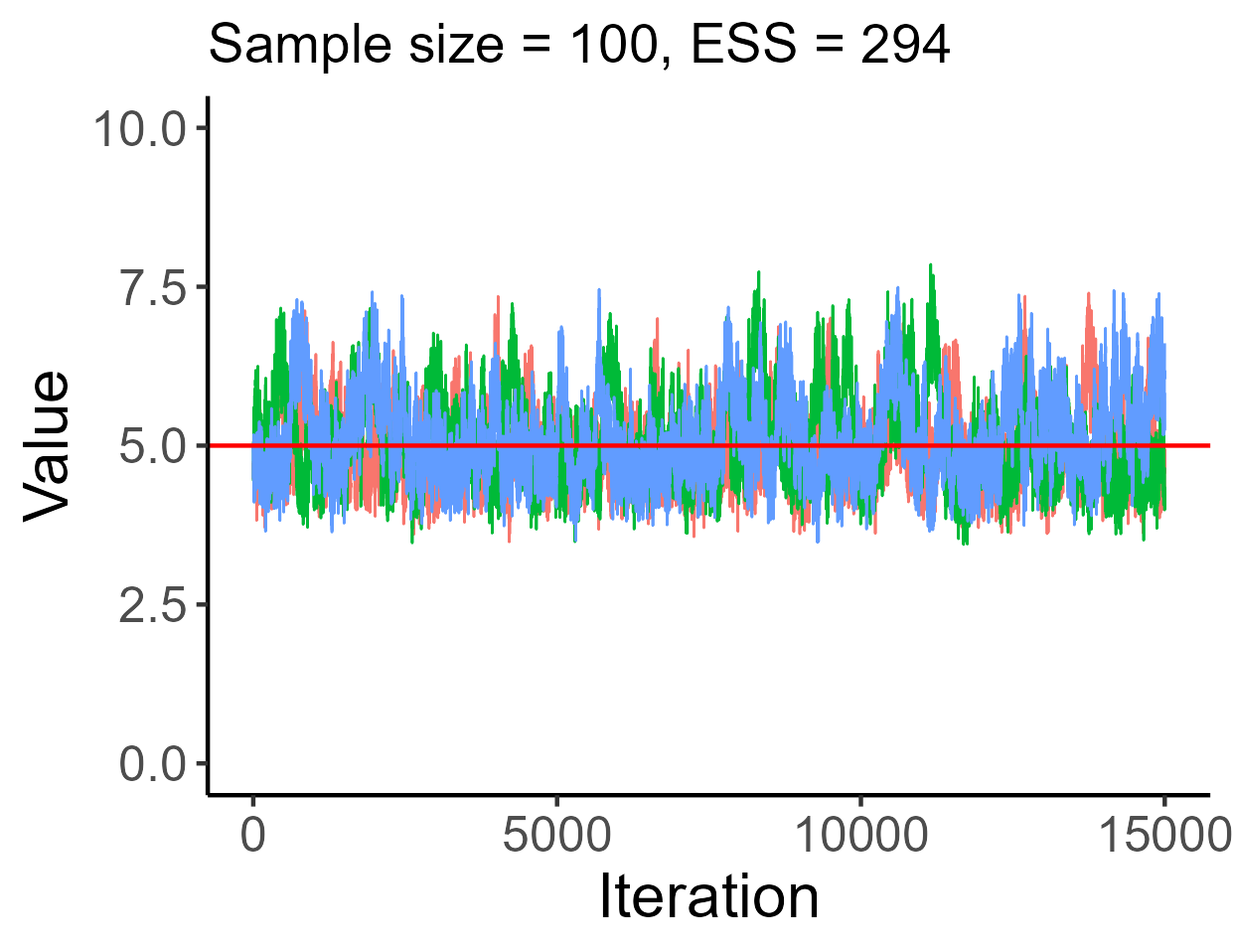}
\includegraphics[width=0.4\linewidth]{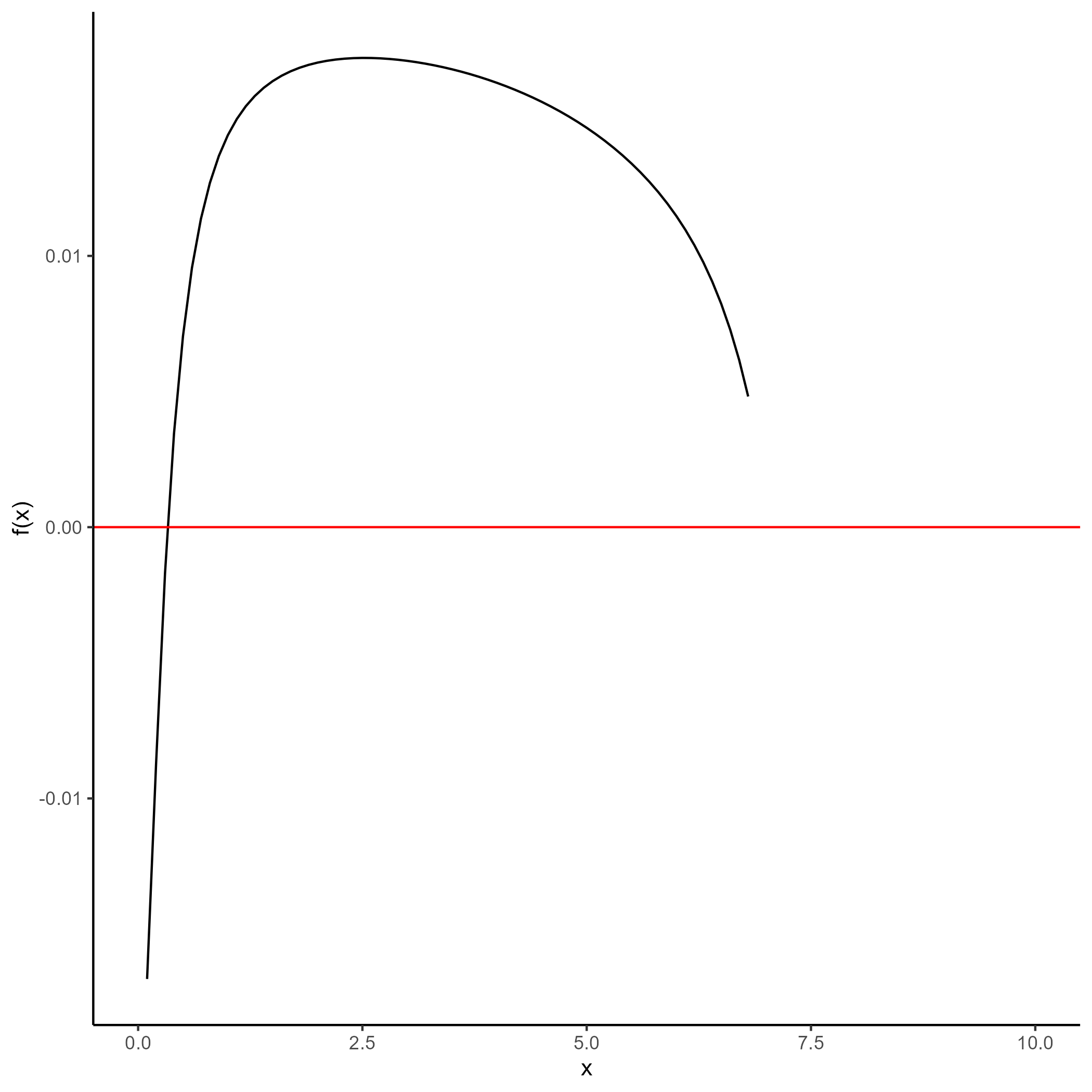}\includegraphics[width=0.4\linewidth]{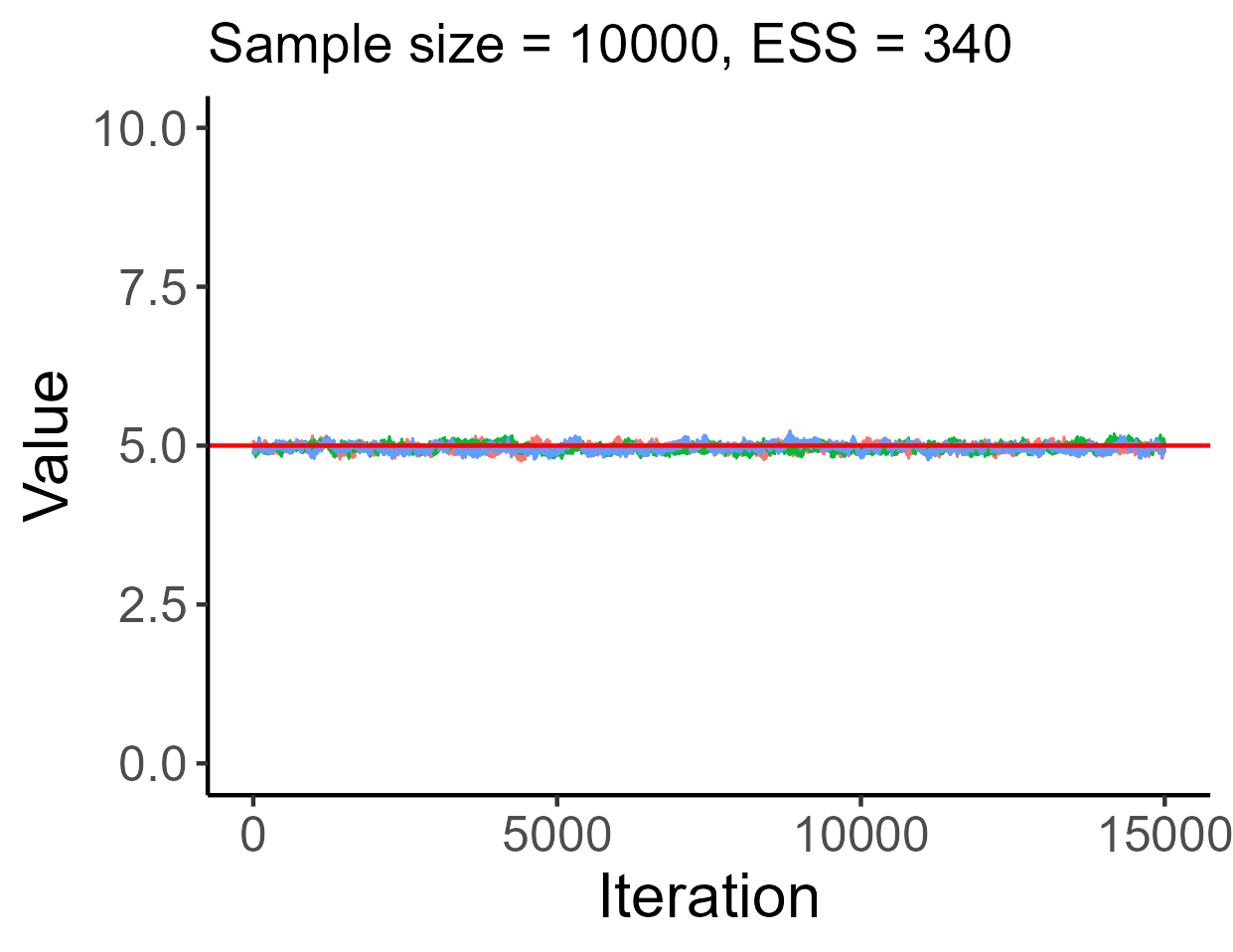}
    \caption{Trace plots and $f(x)$ graphs}
    \label{fig:11:section5-identification}
    The left plots are the $f(x)$ graphs representing the number of solutions for $f(\mu) = 0$. The right plots are the trace plots of $\mu$. The top two plots use the data of \autoref{ex:5:simplecount} with $n = 100$, and the bottom plots use the data with $n = 10000$.
\end{figure}
If we simplify this, the number of solutions $\mu$ in the given equation:
$$f(\mu) = \frac{e^{-\mu}\mu^2}{2} h\left(\frac{c_0}{e^{-\mu}},\frac{c_1}{e^{-\mu} \mu}\right) - c2 = 0$$

To check the computational performance of MCMC and the number of solutions, we first generate the dataset using the same method in \autoref{ex:5:simplecount} and perform MCMC sampling on it. Then, by using estimates $\hat c_i = \frac{\#(Y = i, R = 1)}{n}$, we find a number of the solutions.
In \autoref{fig:11:section5-identification}, when $n = 100$, there is no solution for $f(\mu) = 0$, and the trace plot shows a partial identification. And when $n = 10000$, there is one solution for $f(\mu) = 0$, and the trace plot shows that the model is weakly identified.